\documentclass{nature3}
\usepackage{graphicx}
\usepackage{xcolor}
\usepackage{float}
\usepackage{longtable}
\usepackage{tabularx}
\usepackage{url}
\usepackage{amsmath}
\usepackage{subcaption}
\usepackage{courier}
\usepackage[super]{nth}
\usepackage[labelfont=bf, labelsep=none]{caption}

\usepackage[super,sectionbib]{natbib}
\usepackage[colorlinks=true,citecolor=blue,urlcolor=cyan]{hyperref}
\usepackage{astjnlabbrev}
\usepackage{marvosym}

\usepackage[affil-it]{authblk}

\newcommand{\sep}{$\,\vert\,$}

\usepackage{lineno}

\makeatletter
\def\@maketitle{%
  \newpage
  \begin{center}%
  \let \footnote \thanks
    {\large \@title \par}%
    \vskip -3em
    {\large
      \begin{tabular}[t]{c}%
        \@author
      \end{tabular}\par}%
  \end{center}%
  \par
  \vskip 0.5em}
\makeatother

\title{Highly reflective white clouds on the western dayside of an exo-Neptune}

\author[1,2]{Louis-Philippe Coulombe}
\newcounter{ind}

\stepcounter{ind}\edef\UdeM{\arabic{ind}}
\affil[\arabic{ind}]{Department of Physics, Universit\'{e} de Montr\'{e}al, Montr\'{e}al, QC, Canada}

\stepcounter{ind}\edef\IREx{\arabic{ind}}
\affil[\arabic{ind}]{Trottier Institute for Research on Exoplanets, Montr\'{e}al, QC, Canada}

\stepcounter{ind}\edef\ucla{\arabic{ind}}
\affil[\arabic{ind}]{Department of Earth, Planetary, and Space Sciences, University of California, Los Angeles, Los Angeles, CA, USA}

\stepcounter{ind}\edef\UoO{\arabic{ind}}
\affil[\arabic{ind}]{Department of Physics, University of Ottawa, Ottawa, ON, Canada}

\stepcounter{ind}\edef\McgillP{\arabic{ind}}
\affil[\arabic{ind}]{Department of Physics, McGill University, Montréal, QC, Canada}

\stepcounter{ind}\edef\McgillE{\arabic{ind}}
\affil[\arabic{ind}]{Department of Earth \& Planetary Sciences, McGill University, Montréal, QC, Canada}

\stepcounter{ind}\edef\Azur{\arabic{ind}}
\affil[\arabic{ind}]{Université Côte d'Azur, Observatoire de la Côte d'Azur, CNRS, Laboratoire Lagrange, Nice, France}

\stepcounter{ind}\edef\Oxford{\arabic{ind}}
\affil[\arabic{ind}]{Department of Physics, University of Oxford, Oxford, Oxfordshire, UK}

\stepcounter{ind}\edef\Cornell{\arabic{ind}}
\affil[\arabic{ind}]{Carl Sagan Institute and Department of Astronomy, Cornell University, Ithaca, NY, USA}

\stepcounter{ind}\edef\NRC{\arabic{ind}}
\affil[\arabic{ind}]{NRC Herzberg Astronomy and Astrophysics, Victoria, BC, Canada}

\stepcounter{ind}\edef\UVic{\arabic{ind}}
\affil[\arabic{ind}]{Department of Physics and Astronomy, University of Victoria, Victoria, BC, Canada}

\stepcounter{ind}\edef\JHU{\arabic{ind}}
\affil[\arabic{ind}]{Department of Physics and Astronomy, Johns Hopkins University, Baltimore, MD, USA}

\stepcounter{ind}\edef\umich{\arabic{ind}}
\affil[\arabic{ind}]{Department of Astronomy, University of Michigan, Ann Arbor, MI, USA}

\stepcounter{ind}\edef\bishop{\arabic{ind}}
\affil[\arabic{ind}]{Department of Physics and Astronomy, Bishop's University, Sherbrooke, QC, Canada}

\author[\UdeM,\IREx]{Michael Radica}
\author[\UdeM,\IREx,\ucla]{Björn Benneke}
\author[\IREx,\UoO]{Élyse D'Aoust}
\author[\UdeM,\IREx]{Lisa Dang}
\author[\McgillP,\McgillE,\IREx]{Nicolas B. Cowan}
\author[\Azur]{Vivien Parmentier}
\author[\UdeM,\IREx]{Loïc Albert}
\author[\UdeM,\IREx]{David Lafrenière}
\author[\Oxford,\IREx]{Jake Taylor}
\author[\UdeM,\IREx]{Pierre-Alexis Roy}
\author[\UdeM,\IREx]{Stefan Pelletier}
\author[\UdeM,\IREx]{Romain Allart}
\author[\UdeM,\IREx]{\'Etienne Artigau}
\author[\UdeM,\IREx]{René Doyon}
\author[\Cornell]{Ray Jayawardhana}
\author[\NRC,\UVic]{Doug Johnstone}
\author[\Cornell]{Lisa Kaltenegger}
\author[\Cornell,\JHU]{Adam B. Langeveld}
\author[\umich]{Ryan J. MacDonald}
\author[\bishop,\IREx]{Jason F. Rowe}
\author[\Cornell]{Jake D. Turner}

\DeclareCaptionFormat{cont}{#1 (cont.)#2#3\par}

\begin{document}

\maketitle


\vspace{0mm}
\begin{abstract}
Highly-irradiated gas giant exoplanets are predicted to show circulation patterns dominated by day-to-night heat transport and a spatial distribution of clouds that is driven by advection and local heating. Hot-Jupiters have been extensively studied from broadband phase-curve observations at infrared and optical wavelengths, but spectroscopic observations in the reflected light are rare and the regime of smaller and higher-metallicity ultra-hot planets, such as hot-Neptunes, remains largely unexplored to date.  
Here we present the phase-resolved reflected-light and thermal-emission spectroscopy of the ultra-hot Neptune LTT\,9779b, obtained through observing its full phase-curve from 0.6 to 2.8\,µm with JWST NIRISS/SOSS.
We detect an asymmetric dayside in reflected light (3.1$\sigma$ significance) with highly-reflective white clouds on the western dayside (A = 0.79$\pm$0.15) and a much lower-albedo eastern dayside (A = 0.41$\pm$0.10), resulting in an overall dayside albedo of A = 0.50$\pm$0.07.
The thermal phase curve is symmetric about the substellar point, with a dayside effective temperature of T$_\text{eff,day}$ = 2,260$_{-50}^{+40}$\,K and a cold nightside (T$_\text{eff,night}<$1,330\,K at 3-$\sigma$ confidence), indicative of short radiative timescales.
We propose an atmospheric circulation and cloud distribution regime in which heat is transported eastward from the dayside towards the cold nightside by an equatorial jet, leading to a colder western dayside where temperatures are sufficiently low for the condensation of silicate clouds.
\end{abstract}

We observed the full-orbit phase curve of LTT\,9779b with JWST NIRISS/SOSS \citep{Doyon2023,Albert2023}, tracking the wavelength-dependent modulation of its emitted flux as it rotated more than 360 degrees about its own axis, covering two secondary eclipses and one primary transit. The observations were taken as part of the NIRISS Exploration of the Atmospheric Diversity of Transiting Exoplanets (NEAT) Guaranteed Time Observation Program (GTO 1201; PI D. Lafrenière) on July~\nth{7}, 2022. LTT\,9779b, with its mass of \hbox{$M_\mathrm{p}$ = 29.32$^{+0.78}_{-0.81}$\,$M_{\oplus}$} and radius $R_\mathrm{p}$ = 4.72$\pm$0.23\,$R_{\oplus}$ (ref. \citep{Jenkins2020}), is one of the few known inhabitants of the hot-Neptune desert \citep{Szabo2011,Mazeh2016}. Its short orbit of 0.79 days around its bright \hbox{($J$ = 8.45)} G7V-type host star results in an equilibrium temperature of $T_\mathrm{eq}$ = 1,978$\pm$19\,K, making LTT\,9779b the sole ultra-hot Neptune discovered to date\citep{Jenkins2020}.  
We used the SUBSTRIP256 subarray (256 $\times$ 2048\,pixels) that enables the extraction of the first two SOSS orders; yielding a continuous spectrum from 0.6 -- 2.85\,µm containing the combined light of the host star and the planet throughout all phases of its orbit.
The time series spans 21.94\,h and consists of 4,790 continuous integrations with two groups
and 10.988 seconds of effective integration time, delivering an observing efficiency of 67\%. 
The observations began 1.70\,h before the first secondary eclipse and continued for 0.38\,h after the second one.

\begin{figure}[H]
    \centering
    \includegraphics[width=1.\columnwidth]{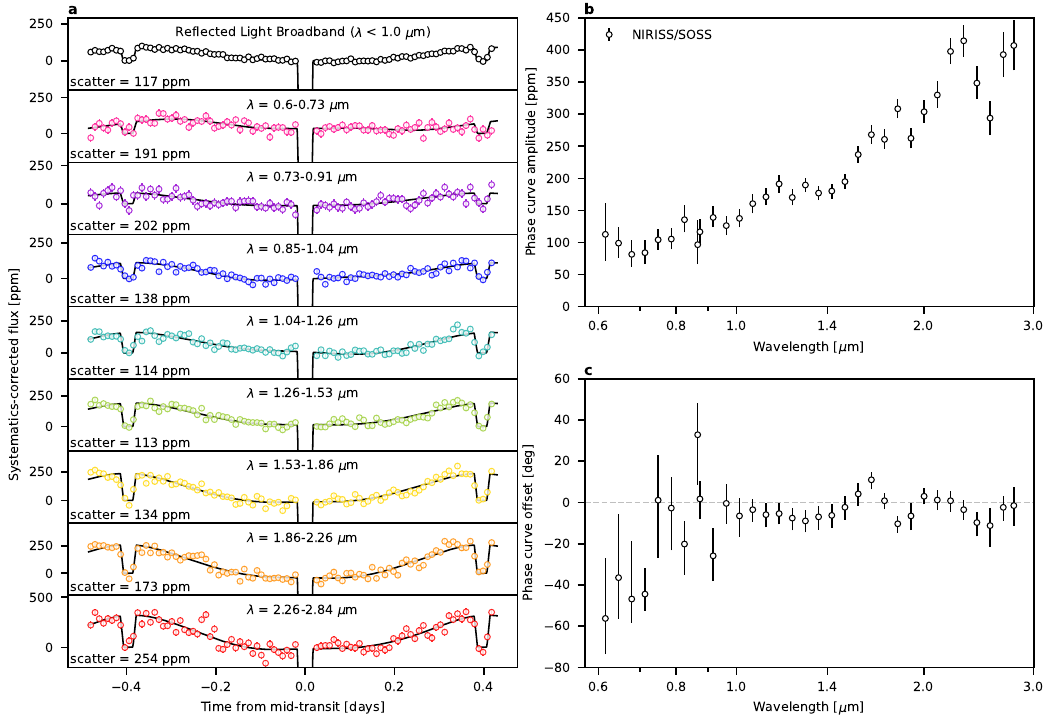}
    \caption{\sep \textbf{White-light and spectroscopic phase curves of LTT\,9779b. a,} Systematics-corrected broadband (black points) and spectroscopic light curves (colored points) along with their best-fit astrophysical models (black lines). The 8 light curves of the second order and 25 light curves of the first order are binned into 2 and 6 light curves, respectively, to display the complete wavelength coverage of NIRISS/SOSS. The data are fitted considering all integrations but are shown with bins of fifty integrations for visual clarity. \textbf{b,} Measured phase curve amplitude, the flux difference between the phase curve maximum and minimum, as a function of wavelength. The phase curve amplitude values, shown with their 1$\sigma$ error bars, are consistent with 100\,ppm below 0.9\,µm, with values monotonically increasing at longer wavelengths as the contribution of thermal emission to the amplitudes becomes important. There is a notable decrease of the phase curve amplitude values in the 1.4\,µm water absorption band. \textbf{c,} Measured phase curve offset, corresponding to the position of the phase curve maximum relative to the phase of mid-eclipse, as a function of wavelength. Positive offset values correspond to a phase curve maximum eastward of the substellar point, where east is the direction of rotation of the planet. The phase curve offset values, shown with their 1$\sigma$ error bars, are consistent with 0$^\circ$ over the near-infrared and decrease to -55$^\circ$ towards optical wavelengths.}
    \label{fig:spec_lcs}
\end{figure}

For this analysis of the full phase curve, the time series observations were reduced using the \texttt{NAMELESS} pipeline following a procedure similar to previously published NIRISS/SOSS datasets \citep{Coulombe2023,Feinstein2023,Radica2023,Lim2023}.
Since the integrations consist of only two groups, we carefully treated cosmic rays and $1/f$ noise as these are the major sources of noise in this observing regime (see Methods; Extended Data Fig. \ref{fig:reduc_steps}).
As part of the analysis verification process, we confirmed that our analysis of the full phase curve presented here, as a side effect, also produces a transmission spectrum consistent with the transit-only light-curve fit presented in ref. \cite{Radica2024}.
We analyzed the extracted light curves by fitting the parameters for the transit, secondary eclipse, phase curve, stellar granulation, and systematics (see Fig. \ref{fig:spec_lcs}a and Extended Data Fig. \ref{fig:wlc_fits}). 
Because of the overlap between the thermal emission of the planet and stellar light reflected by its atmosphere over the NIRISS/SOSS wavelength range, we use an astrophysical model that can produce the phase-resolved signal of a planet with non-uniform albedo and temperature. We model the flux emitted by the planet over its orbit by dividing the planetary surface into six longitudinal slices, fitting for both the thermal emission and albedo of each slice for all spectroscopic light curves (see Methods).
In the systematics model, we account for two sudden changes in the trace morphology caused by tilt events from one of the segments of the primary mirror \citep{Rigby2022,Alderson2023} (see Extended Data Fig. \ref{fig:PCA}), which were identified using the principal component analysis procedure presented in ref. \cite{Coulombe2023}. 
Additionally, we observe the presence of time-correlated variations in the light curves with an amplitude on the order of 100\,ppm and timescales of tens of minutes. This cannot be attributed to the modulation of the planetary emission of LTT\,9779b and
this is instead consistent with the behavior of stellar granulation, which has been previously noted in high-precision \textit{Kepler} observations of bright stars \citep{Gilliland_2011,Kallinger2014, Kallinger2016}. We account for this signal using a Gaussian process with a simple harmonic oscillator kernel that replicates the power spectral density of stellar granulation (see Methods and refs. \citep{Harvey1985,Michel_2009}).

We produced a white-light curve by summing the flux at wavelengths $\lambda<$1\,µm, where the planetary signal is dominated by reflected light (see refs. \cite{Dragomir2020,Crossfield2020,Hoyer2023}). We do not consider wavelengths above 1\,µm for the white light curve as we find that when simultaneously considering reflected light, thermal emission, and a Gaussian process in the fitting, the phase curve model absorbs some of the stellar granulation signal, resulting in unphysical phase curve fits (see Methods).
When fitting the white-light curve, we assumed a circular orbit (zero eccentricity\cite{Jenkins2020}), with the planet rotating about its axis at the same frequency and in the same direction as it revolves around its star. The mid-transit time, $T_0$, as well as the quadratic limb-darkening coefficients, $q_{1,2}$, were allowed to vary freely, and we used Gaussian priors for the period, $P$, semi-major axis, $a/R_\mathrm{s}$, and impact parameter, $b$, based on the precise TESS values\citep{Jenkins2020} (see Methods). 
We obtained the spectroscopic light curves by summing the observed flux at a fixed resolving power of R = 20 (Fig. \ref{fig:spec_lcs}a), yielding a total of 33 wavelength bins (one of which is not considered in the atmospheric analysis, see Methods).
The system parameters were subsequently fixed to their median retrieved values from the white-light curve fit when fitting the spectroscopic light curves.

The phase curve amplitudes and offsets of the phase curve maximum display a clear transition from a reflected light-dominated part of the spectrum to a thermal emission-dominated part (Fig. \ref{fig:spec_lcs}b and c). The phase curve amplitudes show a plateau at 100\,ppm for wavelengths below 0.9\,µm, in agreement with the broadband TESS and CHEOPS observations that found excess flux at optical wavelengths which could not be attributed to thermal emission (see refs.\citep{Dragomir2020,Crossfield2020,Hoyer2023}, also Extended Data Fig. \ref{fig:sec_ecl_spec_ppm}). This excess observed planetary flux can only be explained by reflective clouds, as Rayleigh scattering from the gas opacity would rise sharply at shorter wavelengths \citep{Marley_1999} and result in a symmetric phase curve in the optical \citep{Demory2013}. We also observe a significant dip in the phase curve amplitude around the 1.4\,µm water absorption feature, indicative of a non-inverted atmosphere with temperatures decreasing with altitude on the dayside. A possible explanation for the absence of an inversion is that the optical absorbers assumed to be responsible for thermal inversion in hot Jupiters, such as TiO and VO, have condensed and gravitationally settled on the nightside of LTT 9779b (ref. \cite{Parmentier_2013}).
The phase-curve offset also varies significantly over the NIRISS/SOSS wavelength range. At wavelengths below 1\,µm, where reflected light is dominant, the phase curve maximum is offset to the west and reaches values of -55$^\circ$ at the shortest wavelengths, indicating that more light is reflected from the western dayside (3.1$\sigma$ significance for an asymmetry, see Methods and Extended Data Fig. \ref{fig:Aw_vs_Ae}). The increasingly westward offset at short wavelengths can be explained by longitudinal non-uniformity in the cloud deck across the dayside \citep{Demory2013,Shporer_2015,Parmentier2016,Parmentier_2020}. At near-infrared wavelengths, the phase curve offset is relatively constant around 0$^\circ$, indicating a maximum of the phase curve near the substellar point of the planet. We also observe a decrease of the phase-curve offset inside the water absorption bands, which we interpret as the planet being less efficient at redistributing heat towards the nightside at lower pressures, where the radiative timescales are shorter \citep{Crossfield2020}.

\begin{figure}[H]
    \centering
    \includegraphics[width=1.\columnwidth]{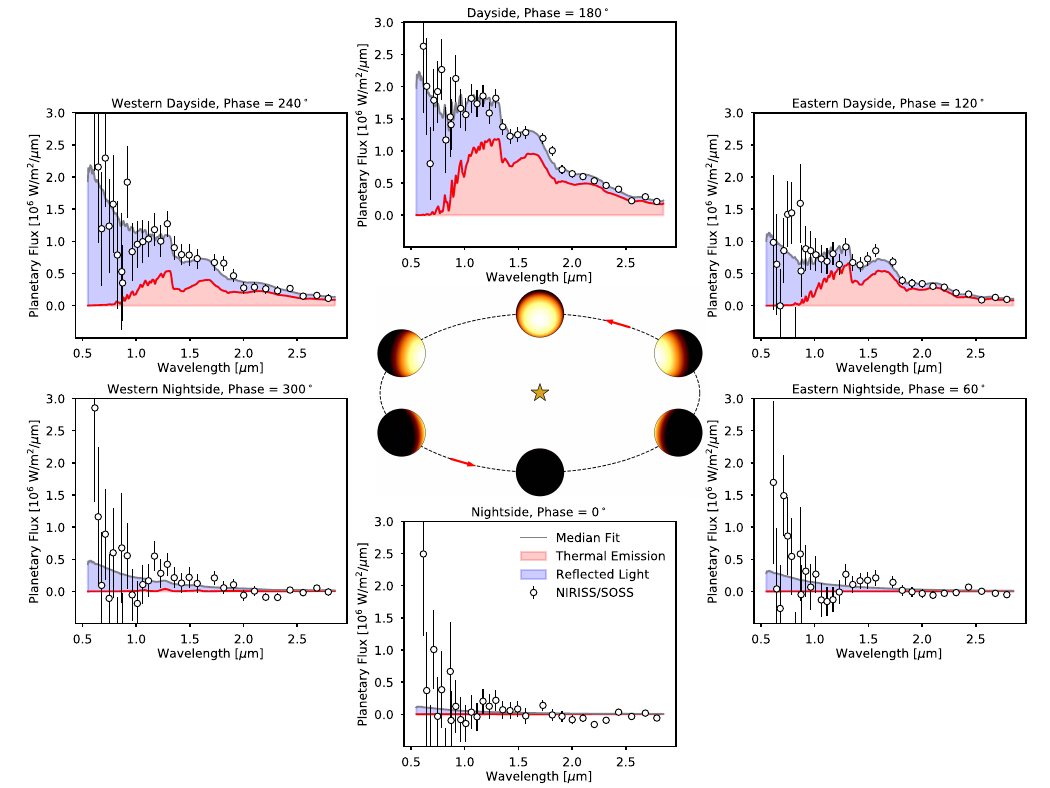}
    \caption{\sep \textbf{Phase-resolved planetary flux of LTT\,9779b.} Measured spectra at six orbital phases and converted from planet-to-star flux ratio values to physical units of flux (black points, shown with 1$\sigma$ error bars) using the PHOENIX stellar spectrum model considered in the atmospheric retrievals (see Methods). The median retrieved planetary spectra from the atmospheric retrievals assuming chemical equilibrium are shown by the gray lines, with the estimated contributions of thermal emission and reflected light from the best-fitting retrieval model shown by the red and blue shaded regions, respectively. In the center of the figure is a schematic of LTT\,9779b as it orbits around its host star at the six orbital phases from which we have extracted the spectra. At phase = 180$^\circ$ the full dayside hemisphere is in view, while at phase = 0$^\circ$ the full nightside is in view. There is a noticeable difference between the eastern and western dayside at short wavelengths. The long-wavelength data, where thermal emission dominates the planetary flux, provide precise constraints on the temperature-pressure profile and, consequently, the amount of thermal emission that is expected at shorter wavelengths (Extended Data Fig. \ref{fig:therm_refl_constraint}). This constraint on the short-wavelength thermal emission enables us to infer the exact amount of flux that is contributed by reflected light.}
    \label{fig:Fp_vs_phase}
\end{figure}

\begin{figure}[H]
    \centering
    \vspace{-5mm}
    \includegraphics[width=1.\columnwidth]{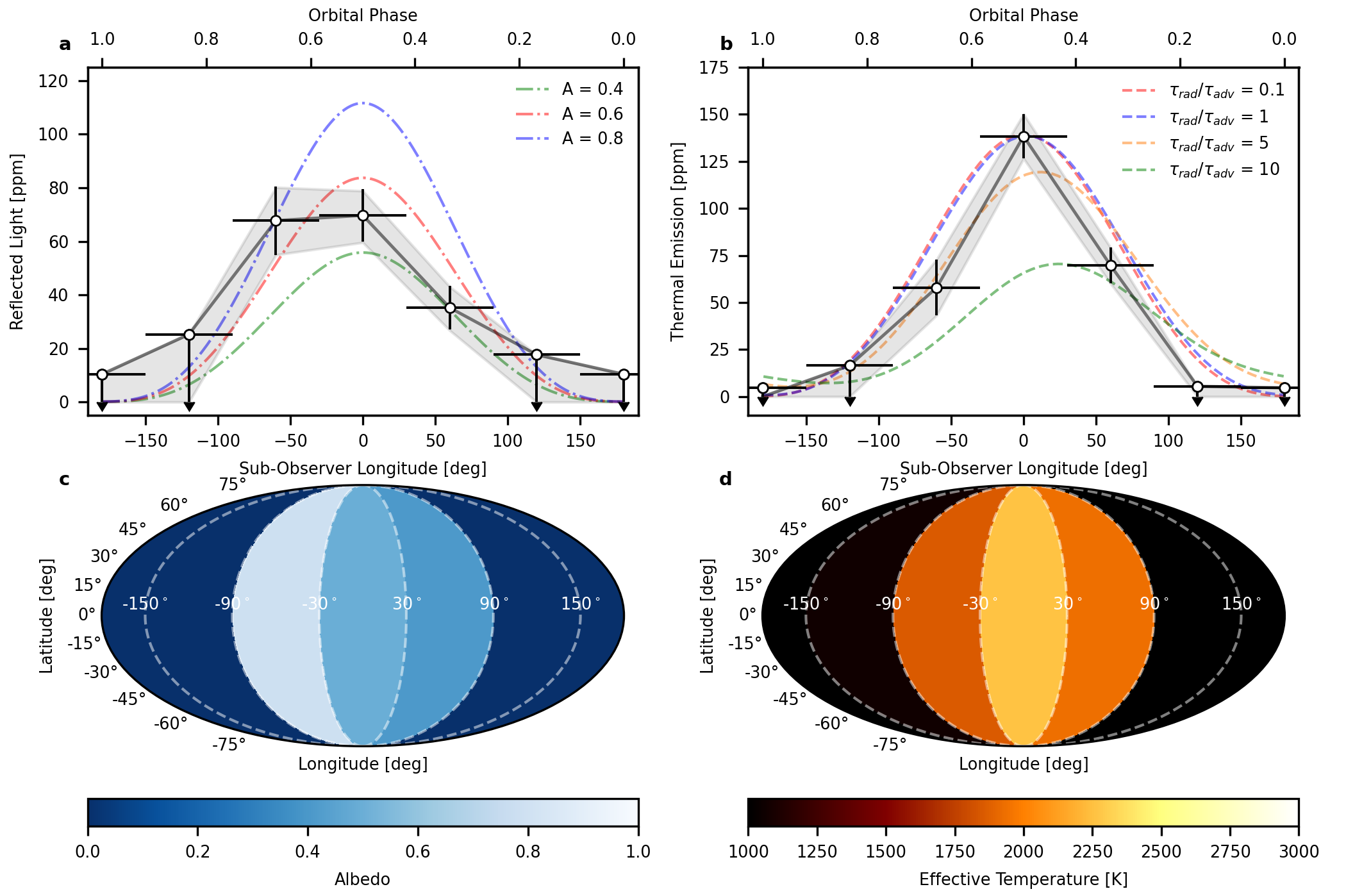}
    \caption{\sep \textbf{Phase-resolved reflected light and thermal components of LTT\,9779b. a,} Reflected light amplitude as a function of the sub-observer longitude -- the longitude at the center of the projected planetary disk for a given orbital phase. The reflected light values are derived from the phase-resolved chemical equilibrium retrievals. Median values (dots, with 1-$\sigma$ uncertainties) and 2-$\sigma$ upper limits (arrows) are shown and also indicated by the black line and shaded region.
    The retrieved reflected light profile shows a strong westward asymmetry, deviating from a Lambertian reflector (dash-dotted green, red, and blue lines for geometric albedo values of 0.4, 0.6, and 0.8, respectively). \textbf{b,} Bandpass-integrated thermal emission for wavelengths 1.0$\leq\lambda\leq$2.8\,µm as a function of sub-observer longitude (see Methods). Median values (dots, with 1-$\sigma$ uncertainties) and 2-$\sigma$ upper limits (arrows) are shown and also indicated by the black line and shaded region.
    The thermal emission shows a maximum of the disk-integrated flux at phase 0.5 with a sharp and symmetric decrease in flux towards the cold nightside. Energy balance models are shown
    by the dashed lines for radiative-to-advective timescale ratio $\tau_\text{rad}/\tau_\text{adv}$ values of 0.1 (red), 1 (blue), 5 (orange), and 10 (green) (see Methods). \textbf{c,} 
    Map where the slices have been assigned the albedo values corresponding to the measured disk-integrated reflected light as a function of orbital phase.
    The western portion of the dayside shows an albedo of 0.79$\pm$0.15, which gradually decreases to 0.41$\pm$0.10 towards the eastern dayside. Past the terminators, the albedo is unconstrained and is shown as being equal to 0.
    \textbf{d,}
    Map where the slices have been assigned the effective temperature values corresponding to the measured disk-integrated thermal emission as a function of orbital phase.
    The effective temperature peaks at  2,260$_{-50}^{+40}$\,K at the substellar point, with the eastern and western dayside having effective temperatures of 1,930$_{-60}^{+60}$\,K and 1,850$_{-110}^{+100}$\,K, respectively.}
    \label{fig:Ag_Teff_vs_phase}
\end{figure}

\begin{figure}[H]
    \centering
    \includegraphics[width=1.\columnwidth]{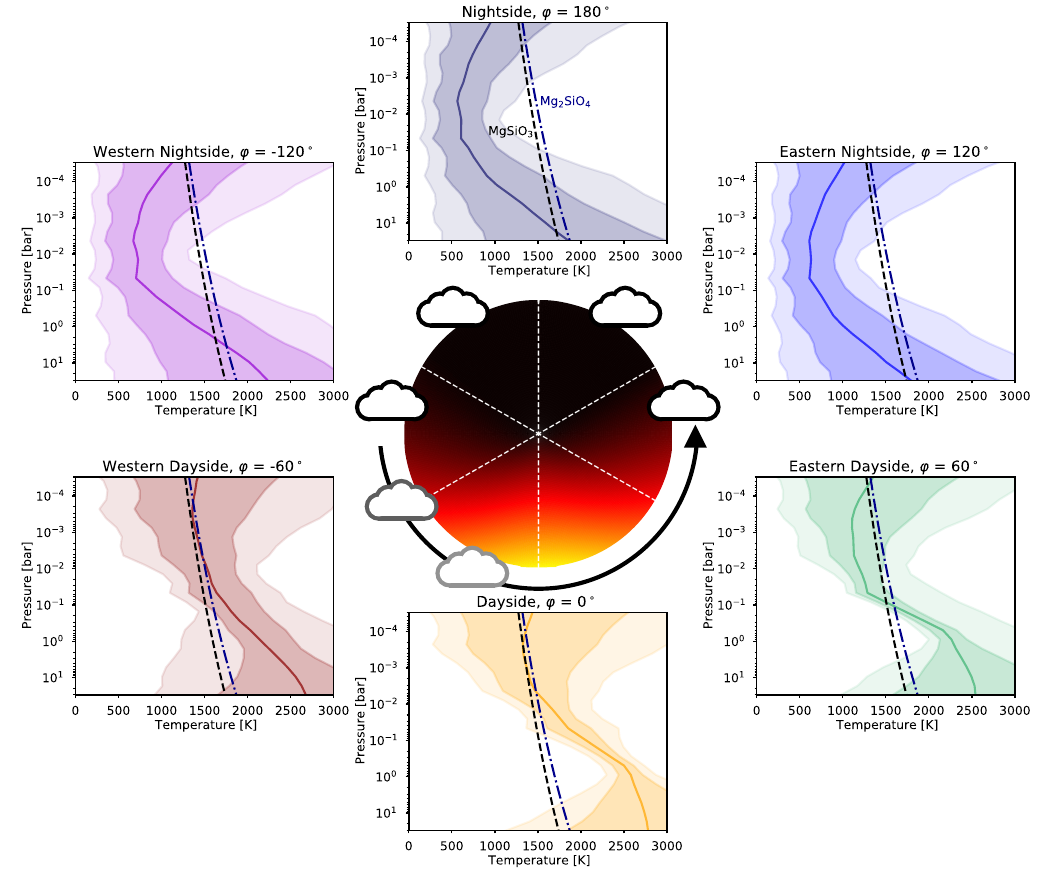}
    \caption{\sep \textbf{Longitudinal transition of LTT\,9779b from cloudy to cloud-free.} Retrieved median (full lines) temperature-pressure profiles for the planetary emission spectra extracted at six orbital phases. Longitude $\varphi$ =  180$^\circ$ corresponds to the nightside whereas $\varphi$ = 0$^\circ$ is the dayside. The 1 and 2-$\sigma$ confidence intervals are shown by the shaded regions. The MgSiO$_3$ (enstatite, dashed black line) and Mg$_2$SiO$_4$ (forsterite, dash-dotted blue line) condensation curves from ref. \cite{Visscher_2010} are shown. These species can condense out of the atmosphere and form clouds when the local temperature is lower than the corresponding condensation curve. In the center of the figure is a schematic depicting the temperature and cloud coverage of LTT\,9779b as seen from above the orbital plane. The white dashed lines indicate the orbital phases analyzed by the six phase-resolved atmospheric retrievals. The black arrow indicates the direction of the equatorial jet caused by the day-night temperature gradient.
    There is an asymmetry in cloud coverage, as heat is advected from the dayside to the nightside by an eastward equatorial jet, leading to a colder western dayside where silicate clouds can form.}
    \label{fig:TP_vs_phase_row}
\end{figure}

Atmospheric retrievals performed on the phase-resolved spectra, extracted at six orbital phases (0$^\circ$, 60$^\circ$, 120$^\circ$, 180$^\circ$, 240$^\circ$, 300$^\circ$) from the phase curve fits, reveal a highly reflective western dayside and a maximum of the thermal emission at the substellar point (Fig. \ref{fig:Fp_vs_phase} and Extended Data Fig. \ref{fig:phase_specs}). 
The atmospheric analysis was done using the SCARLET framework \citep{Benneke2012,Benneke2013,Benneke2015,Benneke2019} considering chemical equilibrium and a free non-parametric temperature profile (see Methods). All three phases covering the dayside of LTT\,9779b (120$^\circ$, 180$^\circ$, and 240$^\circ$) show both significant reflected light at short wavelengths and thermal emission past 1\,µm. Consistently with the observed phase curve offset spectrum, thermal emission is at its maximum at the substellar point, and we observe that the amount of reflected light received from the western dayside is twice that of the eastern dayside. The negligible amount of reflected light observed from the phases covering the nightside (0$^\circ$, 60$^\circ$, and 300$^\circ$) is expected given the small area of the planet that reflects light back towards the observer at these geometries\citep{Cowan_2013}.
With the ability of the atmospheric retrievals to disentangle the contributions of reflected light and thermal emission to the planetary flux (Extended Data Fig. \ref{fig:therm_refl_constraint}), we are able to phase-resolve the albedo and effective temperature distributions of LTT\,9779b (Fig. \ref{fig:Ag_Teff_vs_phase} and Extended Data Fig. \ref{fig:Teff_Ag_all}). We measure a highly reflective dayside for LTT\,9779b, with an albedo of $A$ = 0.50$\pm$0.07, and find that its albedo varies from 0.79$\pm$0.15 to 0.41$\pm$0.10 on the western and eastern daysides, respectively. We also measure a dayside effective temperature of $T_\text{eff,day}$ = 2,260$_{-50}^{+40}$\,K, with a symmetric decrease of 300--400\,K towards the western and eastern daysides. Furthermore, the low thermal emission observed at the three nightside phases provides upper-limits on their effective temperatures, with a nightside effective temperature of $T_\text{eff,night}<1,330$\,K at 3-$\sigma$ confidence. When comparing our measured phase-resolved thermal emission to phase curves produced using the energy transport model presented in ref. \cite{Crossfield2020} (see Methods),
we find that cases where a significant amount of heat is re-emitted from the nightside ($\tau_\mathrm{rad}\gg\tau_\mathrm{adv}$) cannot explain the observed flux distribution and measured near-infrared phase curve offsets (Fig. \ref{fig:Ag_Teff_vs_phase}d).
To first order, the circulation of tidally-locked gas giants, such as hot-Jupiters and hot-Neptunes,
is driven by the difference in irradiation between its permanent day and nightsides, leading to the formation of a superrotating eastward equatorial jet \citep{Showman_2002,Showman_2009,Lewis2010,Komacek_2016}. 
Our measured temperature and albedo distributions of LTT\,9779b are the result of this circulation regime, as the transport of heat eastward of the substellar point produces a colder western dayside where temperatures are sufficiently low for clouds to form (Fig. \ref{fig:TP_vs_phase_row}).
Condensation curves for MgSiO$_3$ and Mg$_2$SiO$_4$, two species that are expected to condense at the equilibrium temperature of LTT\,9779b \citep{Gao_2020,Wakeford2015,Wakeford_2016} and best explain our observed reflected light spectrum (Extended Data Fig. \ref{fig:cloud_models}), show that clouds can indeed form or persist on its nightside and western dayside.
This scenario is consistent with our measured thermal phase curve as well as the 4.5\,µm \textit{Spitzer}/IRAC phase curve measurement of LTT\,9779b (ref. \citep{Crossfield2020}), as nightside clouds and super-solar atmosphere metallicities are both expected to lead to small thermal phase curve offsets and large amplitudes \citep{Lewis2010,Zhang2017,Drummond_2018,Roman2019,Parmentier_2020}. Specifically, the sharp onset of clouds near the terminator is known to produce a thermal phase curve that is more symmetric about the substellar point despite the dayside temperature distribution being asymmetric\citep{Parmentier_2020}. 
Assuming that LTT\,9779b has a super-solar atmospheric metallicity, following the trend observed in solar system planets\citep{Kreidberg_2014,Welbanks_2019}, it would be more prone to the formation of condensates compared to solar-metallicity atmospheres\citep{Visscher_2010}, which is consistent with our measurement of a geometric albedo that is significantly higher than that of hot-Jupiters in the same temperature regime\citep{Wong2021}. This is also consistent with the fact that westward phase curve offsets measured in reflected light with \textit{Kepler} have generally been observed for hot-Jupiters colder than LTT\,9779b ($T_\mathrm{eq}$ = 1400 -- 1800\,K)\citep{Demory2013,Shporer_2015,Parmentier_2020}. This partial coverage of clouds over its dayside, which reflect a certain fraction of the stellar flux, likely affects the energy budget of the planet. We measure a Bond albedo of $A_\mathrm{B}$ = 0.31$\pm$0.06 from our constraints on the dayside and nightside temperatures (see Methods), in contrast with ultra-hot Jupiters which are generally found to absorb practically all incoming radiation\citep{Cowan_2011}.
The presence of clouds over the terminator would also explain the flatness of its transmission spectrum\citep{edwards2023characterising,Radica2024},
indicating that high mean molecular weight alone is not responsible for the lack of observed atmospheric features. Moreover, we detect water absorption features on the dayside of LTT\,9779b (see Methods), further confirming thermal emission spectroscopy as an effective pathway for the atmospheric characterization of cloudy worlds\citep{Kempton_2023}.
Our findings provide crucial insight into the dynamics of smaller highly-irradiated gas giants, a regime that is to date mostly unexplored from a modeling and theoretical standpoint given the scarcity of observations of these planets. In particular, our work shows that the optical-to-infrared wavelength coverage that is enabled by JWST NIRISS/SOSS (0.6--2.8\,µm) and NIRSpec/PRISM (0.5--5.5\,µm) represents a powerful tool to study the interplay of atmospheric circulation and cloud distribution for a wide range of gas giant exoplanets with reflected-light and thermal-emission spectroscopy.

\clearpage


\begin{methods}

\subsection{NIRISS/SOSS Reduction and Spectroscopic Extraction.}\label{sec:extraction}

We use the \texttt{NAMELESS} data reduction pipeline to go from the uncalibrated observations up to the extraction of the spectroscopic light curves. The main steps of this pipeline have been described in detail in refs. \cite{Feinstein2023,Coulombe2023}.
We go through all stage 1 steps of the STScI jwst reduction pipeline \citep{bushouse_2023}, except for the dark subtraction step, as the \texttt{jwst\_niriss\_dark\_0171.fits} reference file shows signs of residual $1/f$ noise. Despite the fact that a few pixels are fully saturated near the peak of the order 1 throughput, we find no resulting bias in our extracted spectra \citep{Radica2024}. We proceed after the ramp-fitting step and apply the world coordinate system, source type, and flat field steps from the jwst pipeline stage 2. After these steps have been performed, we go through a sequence of custom routines to correct for the following sources of noise: bad pixels, background, cosmic rays, and read noise (commonly referred to as $1/f$ noise, where $f$ is frequency).

We begin our data reduction with bad pixel correction, which aims to correct for outliers in the detector images that are present at all integrations. The frames of each integration are divided into windows of 4$\times$4 pixels, for which we compute the two-dimensional second derivative of the pixels it contains. The choice of a 4$\times$4 window size is made as it was found to produce best results in past NIRISS/SOSS analyses \cite{Feinstein2023,Coulombe2023,Lim2023}.
All pixels with second derivative more than $4\sigma$ above the median second derivative of the 4$\times$4 window they are found within are flagged as outliers.
Furthermore, we observe that some bad pixels show a “plus sign” pattern, with pixels adjacent to an outlier pixel also showing count levels that are larger than expected. We flag these by looking at the 3$\times$3 array around the bad pixel and checking if the median in time of the four pixels touching the center pixel is larger than that of the four corner pixels, in which case the touching pixels are flagged. We further flag all pixels showing negative or null count rates. Finally, we create a median outlier map by computing the median frame of the outliers flagged for each integration, i.e., if a pixel is flagged in more than 50\% of all integrations, it is considered a bad pixel. We then go through each integration and assign a new value to all flagged pixels using cubic two-dimensional interpolation computed from the non-flagged pixels with the  \texttt{scipy.griddata} python module. While this method is efficient at correcting outliers that are common to all integrations, a separate algorithm is needed to correct for cosmic rays, which affect a given pixel over a single or a few integrations.

Background correction is performed using the model background made from commissioning observations and provided by STScI in the JDox User Documentation\cite{jdocs}. Similar to what was observed with the NIRISS/SOSS observations of TRAPPIST-1b 
(ref. \citep{Lim2023}), we find that the model background is unable to reproduce the amplitude of the jump -- a break in the background flux happening around spectral column $x = 700$ (ref. \citep{Albert2023}) -- in the observed background. To account for this, we divide the background model into two sections separated by the jump in background flux and scale them separately to the observed background. We note that the jump in background flux is not discrete, with the transition occurring over a dozen pixels, this transition region is scaled using the same factor as the region after the jump in flux, which we find produces a good fit of the observed background. The regions of the detector used to scale the two sections of the model background ($x_1\in[350,550]$, $y_1\in[235,250]$ and $x_2\in[715,750]$, $y_2\in[235,250]$) were chosen because of their distance from the three spectral traces and their lack of contaminants. The scaling factors are computed using the 16$^{\text{th}}$ and 0.18$^{\text{th}}$ percentiles of the scaling distributions for the first and second region of the background, respectively, as we find these values to produce the lowest discrepancy between the regions separated by the jump after background subtraction. Finally, we remove the scaled model background from all integrations.

To detect outliers caused by cosmic rays, we look at the time domain, contrary to bad pixels which are found in the spatial domain. Because our observations only have two groups per integration, the jump detection algorithm from the jwst pipeline is unable to detect and correct for cosmic ray hits, as the algorithm used requires at least three groups per integration as of version v1.8.5. We instead opt to detect and correct for cosmic rays by computing the running median in time for all individual pixels, using a window size of 10 integrations, and bringing all integrations where the count of a pixel is 3$\sigma$ away from the running median to the latter's value. When computing the error value of a pixel, we use the half-width between the 16$^{\text{th}}$ and 84$^{\text{th}}$ quantiles rather than the standard deviation, as this measure is less sensitive to large outliers. To ensure that $1/f$ noise is not picked up by our cosmic ray detection algorithm, we first subtract the median frame from all integrations, scaled using a smoothed version of the white light curve computed by summing pixels $x \in [1200,1800],~ y\in[25,55]$. We subsequently subtract the median column value for each column and integration as a temporary measure to correct for $1/f$ noise. The cosmic ray detection is then performed on the median-subtracted and $1/f$-noise-corrected integrations. If a cosmic ray is detected on a given pixel at some integration, its value is replaced by that of its running median. Once the cosmic rays have been corrected, we add back the median columns and the scaled median integration to all integrations, leaving a more careful treatment of the $1/f$ noise for the following step.

We then finally correct for the $1/f$ noise, which has a significant effect at the integration level for these observations, as they contain only two groups. The method used for this dataset is similar to that of ref. \cite{Coulombe2023} with a few modifications to account for the presence of the third and second spectral orders which were not included in the observations of WASP-18b.
We begin by producing median frames for the three portions of the observations that are separated by the two tilt events observed (see Instrument Systematics section). We then scale all columns of the median frame independently at each integration while fitting for an additive constant that represents the $1/f$ noise such that the chi-square between the observed and scaled columns is minimized. The errors considered in the computation of the chi-square are those returned by the jwst pipeline. Pixels showing non-zero data quality flags have their errors set to $\infty$ (resulting in null weights) such that they are not considered when computing the $1/f$ noise.
For both the first and second spectral orders, We scale each column individually instead of considering a single scaling for the whole trace, as the planetary signal is highly wavelength-dependent.
Scaling the trace uniformly can lead to a significant dilution of the planetary flux and transmission spectra at long wavelengths, similar to what was observed in ref. \cite{Coulombe2023}.
Furthermore, we find that the change in $1/f$ noise along a given column is large enough that assuming a constant value leads to significantly higher scatter in the extracted spectra. Similar to the methods used in the \texttt{transitspectroscopy} \citep{Feinstein2023} and \texttt{JExoRES} pipelines \citep{Holmberg_2023,madhusudhan2023carbonbearing}, we only consider pixels inside a given window centered around the trace to compute the $1/f$ noise values. First, we set the errors of all pixels that are more than 30 pixels away from the center of the first order trace to $\infty$, which means that the scaling values and $1/f$ noise is solely computed from the variation of the first order. Second, we subtract from each column the value of $1/f$ noise computed from the window centered around the first order trace. While this results in frames that are free of $1/f$ striping near the first order, we observe residual striping around the second order from the inter-column variation of the noise. We therefore perform a second iteration of the $1/f$ noise where we mask all pixels that are more than 30 pixels away from the center of the second order trace. We also mask pixels that overlap with the window of the first order trace to ensure that the scaling values computed for the second order are not correlated with the first order. Finally, we subtract the second iteration of $1/f$ noise values from the second order window. We find that this method reduces the scatter by up to 30\% at long wavelengths when compared to using the full column.

Finally, after correction of the noise sources discussed above, we proceed with spectral extraction of the first and second orders. We used a simple aperture extraction, since the level of contamination due to the order overlap is predicted to be negligible for this dataset ($\sim$3\,ppm \citealp{Darveau-Bernier2022, Radica2022}). We chose an aperture diameter of 40\,pixels, which we found to minimize the overall scatter.

\subsection{Light Curve Analysis.}\label{sec:lc_fit}

Our observations cover the complete orbit of LTT\,9779b around its host star, including one transit and two secondary eclipses. Modeling of the light curves therefore requires fitting for the orbital and physical parameters of the system, as well as the flux modulation throughout the orbit of the planet coming from the variation in its thermal emission and reflected light as it rotates around itself. We also observe time-correlated noise introduced through variations in the instrument as well as stellar granulation, which we correct for while fitting the astrophysical model. Our model $f(t)$ is therefore described as the combination of these various effects:

\begin{equation}\label{eq:astro_model}
f(t) = F(t) \cdot S_\mathrm{instr}(t) + G(t). 
\end{equation}

\noindent 
An in-depth description of the astrophysical $F$, systematics $S_\mathrm{instr}$, and stellar granulation $G$ models used to fit the light curves is given in the subsections below.

\noindent 
\textbf{Astrophysical Model}
Our astrophysical model is designed to simultaneously model the transit, secondary eclipse, and phase curve flux modulation. The star-normalized system flux as a function of time and wavelength $F(t,\lambda)$ is described as 

\begin{equation}\label{eq:astro_model}
F(t,\lambda) = \mathcal{T}(t,\Theta) + \frac{F_\mathrm{p}(t,\lambda)}{F_\mathrm{s}(t,\lambda)} \mathcal{E}(t,\Theta),
\end{equation}

\noindent 
where $\mathcal{T}$ and $\mathcal{E}$ are the transit and secondary eclipse functions, respectively, which are dependent on the system parameters $\Theta$. The flux modulation outside the secondary eclipse is determined by the time and wavelength-dependent ratio of the planetary $F_\mathrm{p}$ and stellar $F_\mathrm{s}$ fluxes. The transit $\mathcal{T}$ and eclipse $\mathcal{E}$ functions are of unity outside of transit and eclipse, respectively, with the value of the transit function going down to the fraction of light emitted by the unocculted portion of the star and that of the eclipse function reaching zero in full eclipse. The parameters $\Theta$ considered for these functions are the planet-to-star radius ratio $R_\mathrm{p}/R_\mathrm{s}$, time of mid-transit $T_0$, orbital period $P$, semi-major axis $a/R_\mathrm{s}$, impact parameter $b$, and quadratic limb-darkening parameters [$q_1,q_2$] using the parametrization of ref.  \citep{Kipping_2013}.

Because the amount of thermal emission and reflected light emitted by LTT\,9779b is not expected to be constant with orbital phases, we must adopt a parametrization to describe the shape of our observed phase curve. We begin by assuming the stellar flux to be constant in the astrophysical model $F_\mathrm{s}(t,\lambda) = \bar{F}_\mathrm{s}(\lambda)$, with any stellar variation being handled in the systematics model.
We do not consider the Doppler boosting and ellipsoidal variation effects in our astrophysical model, as the amplitudes of these effects are proportional to the planet-to-star mass ratio\cite{Shporer_2017}, which is much smaller in the case of the LTT 9779 system compared to typical hot Jupiter systems. Nevertheless, we estimate the amplitudes of the Doppler boosting and ellipsoidal effects to confirm that their impact on the phase curve is negligible.
Following the analytical formulations of the Doppler boosting amplitudes of refs. \cite{Shporer_2017,Shporer2019,Addison_2021}, we find the expected amplitude to be $A_\mathrm{Boost} = $ 0.5\,ppm at 0.74\,µm, which is the effective wavelength of NIRISS/SOSS spectral order 2 for these observations and where this effect is expected to be strongest. As for the ellipsoidal variation, we use the limb- and gravity-darkening coefficients from ExoCTK \citep{matthew_bourque_2021} and ref. \cite{Claret2017}, respectively, from which we derive an expected ellipsoidal variation of $A_\mathrm{Ellip} =$3.3\,ppm at 0.74\,µm. To model the planetary flux, we use a slice model \citep{Knutson2007,Cowan2008}, where we divide the planetary surface into 6 longitudinal slices of width 60$^\circ$. For each slice, we freely fit a thermal emission $T_n$ and albedo $A_n$ value, which we multiply to the thermal emission and reflected light kernels described in ref. \cite{Cowan_2013} to obtain the total flux $F_\mathrm{p}$ from the planet at a given time:

\begin{equation}\label{eq:astro_model}
F_\mathrm{p}(t,\lambda) = \sum_{n=1}^{N=6} \left[ A_n(\lambda) \cdot \mathcal{K}_{\text{refl},n}(t) + T_n(\lambda) \cdot \mathcal{K}_{\text{therm},n}(t)\right],
\end{equation}

\noindent 
where the reflected light $\mathcal{K}_{\text{refl},n}$ and thermal $\mathcal{K}_{\text{therm},n}$ kernels for a given slice $n$ are described by the integral of the visibility $V$ and illumination $I$ functions over its solid angle $\Omega_n$:

\begin{equation}\label{eq:kernels}
\mathcal{K}_{\text{refl},n} = \frac{R_p}{a\pi}\int_{\Omega_n} V(\theta,\varphi,t) I(\theta,\varphi)  d\Omega ~ ; ~ \mathcal{K}_{\text{therm},n} = \frac{1}{\pi}\int_{\Omega_n} V(\theta,\varphi,t) d\Omega.
\end{equation}

\noindent 
The planetary longitude $\varphi \in [-\pi,\pi]$ and latitude $\theta \in [-\pi/2,\pi/2]$ are defined such that the sub-stellar position is at coordinates ($\varphi=0,~\theta = 0$). As in ref. \cite{Cowan_2013}, we consider a diffusely reflecting (Lambertian) surface for the reflected light kernel. The visibility function corresponds to the cosine of the angle from the sub-observer position ($\varphi_o(t) = -2\pi (t-T_\mathrm{sec})/P,~\theta_o = i-\pi/2$), where $T_\mathrm{sec}$ is the time of secondary eclipse and $i$ is the orbital inclination, whereas the illumination function is the cosine of the angle from the sub-stellar position ($\varphi_s=0,~\theta_s = 0$):

\begin{equation}\label{eq:astro_model}
V(\theta,\varphi,t) = \max{[\cos\theta\cos\theta_o\cos(\varphi-\varphi_o(t))+\sin\theta\sin\theta_o,0]}
\end{equation}

\begin{equation}\label{eq:astro_model}
I(\theta,\varphi) = \max{[\cos\theta\cos\theta_s\cos(\varphi-\varphi_s)+\sin\theta\sin\theta_s,0]}. 
\end{equation}

\noindent 
We numerically solve for the reflected light and thermal emission kernels at each time step for an initial grid resolution of 180 by 360\,pixels to ensure sufficient numerical precision before binning them to the resolution at which the slices are fit (3 by 6 pixels).
Because of the computational cost of numerically integrating $V$ and $I$ at all time steps for a grid resolution of 180 by 360 pixels, they are computed once at the beginning of the light curve fitting for a given set of orbital parameters and kept fixed afterward, as we do not expect variations of the orbital parameters within their probability distribution to significantly affect the shape of the visibility and illumination functions.
The values of the orbital parameters for which $V$ and $I$ are computed at the white-light curve fitting stage are $T_0$ = 59767.54940\,BJD, $P$ = 0.792052\,days, $a/R_s$ = 3.877, and $b$ = 0.912. The values of $V$ and $I$ are then computed using the median retrieved values of the orbital parameters from the white-light curve fit at the spectroscopic light curve fitting stage.

\noindent 
\textbf{Instrument Systematics}
The raw white-light curve considering wavelengths below 1\,µm shows a long-term slope, possibly caused by instrument systematics or stellar variability, which we correct for using a linear-in-time systematics model. After applying a temporal principal component analysis (PCA) directly to the corrected integrations, following the procedure shown in the analysis of the WASP-18b NIRISS/SOSS observations \citep{Coulombe2023}, we find that two tilt events occurred over the phase curve at the 636$^\text{th}$ and 4615$^\text{th}$ integrations. We also observe variations in components analogous to the full width at half maximum (FWHM) and position in the dispersion direction of the trace, similar to those that were observed for WASP-18b. We correct for the two tilt events, which both occur over a single integration and lead to a sudden change in the flux level, by fitting for a flux offset in the light curves at the integrations mentioned above.
We further correct for the trace morphology variations by linearly detrending against the eigenvalues of the first three principal components. However, these components show long-term trends as well as offsets at the integrations where the tilt events occur (see Extended Data Fig. \ref{fig:PCA}), which we do not want to introduce in our light curves. We thus opt to detrend against “cleaned” eigenvalues, which have been divided into three segments around the two tilt events and subsequently median-subtracted to remove offsets. We further subtract a running median with a window size of 100 integrations from the eigenvalues to remove any long-term trend.
Our full systematics model $S_\mathrm{instr}$ is then defined as the sum of the linear slope, the jumps, and the linear detrending against our “cleaned” eigenvalues:

\begin{equation}\label{eq:astro_model}
S_{instr}(t) = a + b\cdot(t-t_0) + c \lambda_{\text{PCA}_1} + d \lambda_{\text{PCA}_2} + e \lambda_{\text{PCA}_3} + j_1\Theta(t-t_{t,1}) + j_2\Theta(t-t_{t,2}), 
\end{equation}

\noindent 
where $a$ is the normalization parameter, $b$ the linear slope, $c$-$d$-$e$ the amplitudes of the principal component eigenvalues, and $j_1$-$j_2$ the amplitudes of the first and second tilt event jumps occurring at times $t_{t,1}$ and $t_{t,2}$ respectively. The jumps are modeled using the Heaviside function $\Theta(x)$, where $\Theta(x<0) = 0 $ and $\Theta(x\geq0)=1$.


Despite detrending against the observed trace variations throughout the observations, we observe leftover time-correlated signal that can not be explained by instrumental effects nor modulation of the flux received from LTT\,9779b. We find this remaining signal is best explained by stellar granulation, which requires a treatment that is separate to that of the instrument systematics.

\noindent 
\textbf{Stellar Granulation}
Stellar activity and granulation are known to be a limiting factor of the precision at which we can retrieve astrophysical properties from transiting exoplanet observations \citep{Grunblatt2017,Barros_2020,Chiavassa_2017,bruno2021stellar,Sarkar_2018}. Stellar activity is commonly used to refer to the observed photometric variation of a star due to the movement of star spots over its projected surface as it rotates. Given the $\sim$45-day rotation period of LTT\,9779 (ref. \citep{Jenkins2020}), we consider that this effect can be handled with low-order polynomials systematics models over the timescale of our observations. In the case of granulation noise, the observed modulation in the brightness of the star is caused by turbulent convection bringing up hotter material from deeper layers to the photosphere \citep{Barros_2020,Pereira_2019}. This signal operates on multiple timescales ranging from minutes to months depending on the size of the granulation cells. Because the stellar type of LTT 9779 (G7V; $T_\mathrm{eff}$ = 5445\,K, $\log g$ = 4.43, [Fe/H] = 0.25)\citep{Jenkins2020} is close to that of the Sun (G2V), we expect these objects to behave similarly in terms of their granulation signal. Following the scaling relations of ref.\citep{Gilliland_2011} (eqs. 5 and 6), we find an expected granulation signal amplitude of 70\,ppm with a timescale of 3\,minutes, which we should be sensitive to considering the photometric precision of our observations.

Given the stochastic nature of granulation noise, it is virtually impossible to directly model its effect on the light curves. However, a Gaussian Process (GP) using a simple harmonic oscillator (SHO) kernel can effectively reproduce stellar granulation noise as they both have the same power spectral density (PSD) functional form \citep{Harvey1985,Michel_2009}. 
For our GP model, we use the \texttt{celerite} open-source python package \citep{Foreman-Mackey2017} whose time complexity scales as $\mathcal{O}(N)$ ($N$ is the number of data points) compared to standard models scaling as $\mathcal{O}(N^3)$, significantly reducing computation time for our light curve fit which has a significant number of integrations ($N$ = 4790). This improvement in time complexity with \texttt{celerite} is enabled by expressing the covariance function as a mixture of complex exponentials, reducing the amount of operations needed to compute the inverse of the covariance matrix.
The SHO kernel of \texttt{celerite} takes as inputs for its PSD, $S(\omega)$, the signal amplitude, $S_0$, characteristic frequency, $\omega_0$, and quality factor, $Q$:

\begin{equation}\label{eq:astro_model}
S(\omega) = \sqrt{\frac{2}{\pi}}\frac{S_0 \omega_0^4}{(\omega^2 - \omega_0^2)^2 + \omega_0^2\omega^2/Q^2}.
\end{equation}

\noindent 
To reproduce the behavior of stellar granulation, the quality factor is set to $Q = 1/\sqrt{2}$ (ref. \citep{Foreman-Mackey2017}) meaning that only $S_0$ and $\omega_0$ are to be fitted in the light curve. To make the choice of priors considered for the GP parameters more intuitive, we instead fit for the granulation amplitude $a_{\text{gran}} = \sqrt{{\sqrt{2}}S_0\omega_0}$ and timescale $\tau_\text{gran} = 2\pi/\omega_0$ (ref. \citep{Pereira_2019}) which are subsequently converted back to $S_0$ and $\omega_0$ before being passed to \texttt{celerite}.

\noindent 
\textbf{Light Curve Fitting.}
We divide the light curve fitting process in two steps: i) the white-light curve fit and ii) the spectroscopic light curve fits. When simultaneously fitting for a phase curve that accounts for reflected light, thermal emission, and a GP at the white-light curve stage, we find that a long dip in the observed flux that occurs during and after the transit (see Extended Data Fig. \ref{fig:wlc_fits}) leads the fit to prefer negative planetary flux values after transit. To avoid our phase curve fit reaching a nonphysical solution, we limit our parameter space by fitting the GP to the reflected light broadband light curve ($\lambda\leq$1.0\,µm), for which we do not fit the thermal component of our slice model. Because the reflected light component of the astrophysical model can only produce low and positive flux values around the transit, this ensures that the GP accounts for the dip in flux that would otherwise lead to negative flux values.

We opt to fit our white-light curve assuming a circular orbit ($e$ = 0) and considering Gaussian priors based on the measurements presented in ref. \cite{Jenkins2020} for the orbital period, semi-major axis, and impact parameter. This results in a fit where we keep free four orbital parameters: the time of mid-transit $T_0$ ($\mathcal{U}[59767.5,59767.6]$ BJD), orbital period $P$ ($\mathcal{N}[0.7920520,0.0000093^2]$\,days), semi-major axis $a/R_\mathrm{s}$ ($\mathcal{N}[3.877,0.091^2]$), and impact parameter $b$ ($\mathcal{N}[0.912,0.05^2]$); three transit parameters: the planet-to-star radius ratio $R_\mathrm{p}/R_\mathrm{s}$ ($\mathcal{U}[0.01,0.1]$) and the quadratic limb-darkening coefficients [$q_{1}$,$q_{2}$] ($\mathcal{U}[0,1]$); six phase curve parameters: the albedo value of each slice $A_{1-6}$ $\mathcal{U}[0,3/2]$); seven systematics model parameters: the normalization parameter $a$ ($\mathcal{U}[-10^9,10^9]$), linear slope $b$ ($\mathcal{U}[-10^9,10^9]$), amplitudes of the eigenvalues $c$-$d$-$e$ ($\mathcal{U}[-10^9,10^9]$), and jump amplitudes $j_1$-$j_2$ ($\mathcal{U}[-1000,1000]$\,ppm); one scatter parameter: $\sigma$ ($\mathcal{U}[50,10000]$\,ppm); as well as two GP parameters: the granulation amplitude $\log_{10} a_{\text{gran}}$ ($\mathcal{U}[1,2.7]$\,$\log_{10}$[ppm]) and timescale $\log_{10} \tau_\text{gran}$ ($\mathcal{U}[0,2]$\,$\log_{10}$[minutes]), for a total of 23 free parameters. The upper limit of the prior for the albedo values is set to 3/2 to allow our reflected light model to produce geometric albedo values $A_g\in[0,1]$, as a fully-reflective Lambertian sphere ($A_{1-6}$ = 1) corresponds to a geometric albedo of 2/3. We explore the parameter space with the affine-invariant Markov chain Monte Carlo Ensemble sampler \texttt{emcee} \citep{Foreman_Mackey_2013} using 4 walkers per free parameter (92 walkers total) and iterating for 50,000 steps. We run an initial fit with 12,500 steps (25\% of the final fit amount) from which we use the maximum probability parameters as the initial position for the final 50,000 steps. Finally, we discard the first 30,000 steps (60\% of the total steps) as burn-in to ensure our posteriors contain only walkers that have converged. From the white-light curve fit, we constrain the stellar granulation amplitude and timescale to 93$\pm$6\,ppm and 21$_{-2}^{+3}$\,minutes, respectively.
The measured granulation amplitude value is roughly consistent with the expected value of $a_\text{gran}$ = 70\,ppm for a star of the type of LTT 9779. As for the granulation timescale, our measurement is consistent within an order of magnitude with the expected value of 3\,minutes.
One possible explanation for the difference between the measured and expected timescales is that the observed stellar granulation signal is a convolution of granulation, supergranulation, and mesogranulation, which all occur on distinct timescales ranging from minutes to hours and might not be well represented by a single timescale.

We proceed with the spectroscopic light curves, binned at a fixed resolving power of R = 20, fits by fixing the orbital parameters to their median retrieved values from the white light curve fit ($T_0$ = 59767.54942$\pm$0.00022 BJD, $P$ = 0.7920523$\pm$0.0000093\,days, $a/R_\mathrm{s}$ = 3.854$\pm$0.082, and $b$ = 0.9061$\pm$0.0080). Because the shape of the stellar granulation noise is expected to be constant as a function of wavelength, with only its amplitude that could show variation due to the changing contrast between the gas cells and the surrounding photosphere, we use the maximum probability GP model from the white-light curve fit and scale it for each spectral bin using a parameter $m_{\text{gran}}$ ($\mathcal{U}[-10^9,10^9]$). When fitting the spectroscopic bins without accounting for stellar granulation, we find that the structure of the noise is mostly achromatic, justifying our choice to scale the white-light GP model. This results in a fit for each spectroscopic bin where we keep free three transit parameters: $R_\mathrm{p}/R_\mathrm{s}$ and [$q_1$,$q_2$]; twelve phase-curve parameters: the albedo $A_{1-6}$ and thermal emission $T_{1-6}$ ($\mathcal{U}[-1000,1000]$\,ppm) values of each slice; seven systematics parameters: $a$, $b$, $c$-$d$-$e$, and $j_1$-$j_2$; one scatter parameter: $\sigma$; as well as the GP scaling parameter: $m_{\text{gran}}$, for a total of 24 free parameters. We use the same fitting procedure as for the white-light curve. We fit the first spectroscopic bin (lowest wavelength) of each spectral order by iterating for 50,000 steps, then passing each subsequent bin the best-fit parameters of the previous bin as the initial position for the fit and iterating for 12,500 steps only as they are started close to their best-fit solution. As for the white-light curve fit, each spectroscopic bin fit is initially run for 25\% of the steps of the final fit and re-initialized at its best-fit position. The posterior distributions of the fit parameters are obtained after discarding the first 60\% steps. From these fits, we extract our six phase-resolved emission spectra as the median and 1-$\sigma$ confidence interval of the $F_\mathrm{p}(t,\lambda)$ samples (eq. \ref{eq:astro_model}) at orbital phases [0$^\circ$, 60$^\circ$, 120$^\circ$, 180$^\circ$, 240$^\circ$, 300$^\circ$].
Upon visual inspection of the extracted phase-resolved spectra, we find that the wavelength bin at 1.65\,µm is significantly lower than the surrounding wavelength bins at all orbitals phases, possibly due to the presence of an uncorrected bad pixel diluting the planetary flux. We thus opt to not consider this wavelength bin at the atmospheric retrieval stage.
The phase curve amplitude and offset values presented in Fig. \ref{fig:spec_lcs} are also computed from the samples of the phase curve models. 

We also quantify the significance of our measurement of an asymmetry between the western and eastern dayside in reflected light at the light curve level. To do this, we produce a light curve where the flux from all wavelengths below 0.85\,µm, where
the measured westward offsets are most important (Fig. \ref{fig:spec_lcs}), is summed. We then follow the same fitting methodology as for the spectroscopic light curves, this time assuming that all the planetary flux is in the form of reflected light at these wavelengths (see Fig. \ref{fig:therm_refl_constraint}) and fitting for six albedo slices. The resulting phase curve is shown in Extended Data Fig. \ref{fig:Aw_vs_Ae}a and shows a westward phase curve offset as well as an amplitude of approximately 100\,ppm, consistent with the results from our spectroscopic light curve fits (Fig. \ref{fig:spec_lcs}). The constraint on the difference between the reflected light flux from the western dayside (phase =  60$^\circ$) and eastern dayside (phase =  -60$^\circ$) is shown in Extended Data Fig. \ref{fig:Aw_vs_Ae}c. We measure a western dayside flux that is larger than that of the eastern dayside at more than 3-$\sigma$ significance. Precisely, we measure that 99.8\% of the posterior probability samples are above 0, corresponding to a p-value of 0.002 and a significance of 3.1$\sigma$. We repeat a similar procedure on a light curve where we have summed the flux from all wavelengths above 1.9\,µm (at which point thermal emission contributes to more than 70\% of the total planetary flux, see Extended Data Fig. \ref{fig:therm_refl_constraint}), for which we assume that the planetary flux can be modelled with thermal emission alone and thus fit it using six thermal emission slices. As shown in Extended Data Fig. \ref{fig:Aw_vs_Ae}b and d, we find that the resulting thermal emission phase curve is symmetric about the secondary eclipse and that the thermal emission measurements from the western and eastern dayside are consistent within 1$\sigma$. 

\subsection{Atmospheric Analysis.}\label{sec:atm_retrieval}

We perform atmospheric retrievals on our measured phase-resolved planetary flux spectra assuming both chemical equilibrium and free chemistry for the thermal emission, as well as an albedo value for the reflected light.
Analysis of the phase-resolved spectra is performed using the atmospheric retrieval SCARLET framework \citep{Benneke2012,Benneke2013,Benneke2015,Benneke2019}. The model thermal emission spectra are converted from physical units to values of $F_\mathrm{p}/F_\mathrm{s}$ using a PHOENIX stellar spectrum considering previously published physical parameters\cite{Jenkins2020} ($T_\mathrm{eff}$ = 5445\,K, $\log g$ = 4.43, [Fe/H] = 0.25). We compare the model stellar spectrum to our extracted and absolute-flux-calibrated stellar spectrum following the methodology presented in ref.\citep{Coulombe2023} and find a that they are in good agreement. The SCARLET forward model computes the emergent disk-integrated thermal emission for a given set of molecular abundances, temperature structure, and cloud properties. The forward model is then coupled to emcee \citep{Foreman_Mackey_2013} to constrain the atmospheric properties.

For the equilibrium chemistry retrievals, we consider the following species: H$_2$, H, H$^-$ (refs.  \citep{Bell1987,John1988}), He, Na, K (refs.  \citep{vald1995,Burrows_2003}), H$_2$O (ref. \citep{Polyansky_2018}), OH (ref. \citep{Rothman2010}), CO (ref. \citep{Rothman2010}), CO$_2$ (ref.  \citep{Rothman2010}), CH$_4$ (ref. \citep{Yurchenko_2014}), NH$_3$ (ref. \citep{Coles_2019}), HCN (ref. \citep{Barber_2013}), TiO (ref. \citep{McKemmish2019}), VO (ref. \citep{McKemmish2016}), and FeH (ref. \citep{Wende_2010}).
The abundances of these species are interpolated in temperature and pressure using a grid of chemical equilibrium abundances from FastChem2 \citep{Stock_2022}, which includes the effects of thermal dissociation for all the species included in the model. These abundances also vary with the atmospheric metallicity, [M/H] ($\mathcal{U}[-1,3]$), and carbon-to-oxygen ratio, C/O ($\mathcal{U}[0,1]$), which are considered as free parameters in the retrieval. We fit for a cloud top pressure $\log$P$_\text{cloud}$ ($\mathcal{U}[-8,2]$\,$\log[\text{bar}]$), where the temperature structure at pressures larger than P$_\text{cloud}$ are set to an isotherm. The reflected light is handled in the atmospheric retrievals by fitting for an achromatic albedo value $A_0$ ($\mathcal{U}[0,1]$), which contributes to the simulated planetary flux spectrum as F$_\mathrm{p}(\lambda)$/F$_\mathrm{s}(\lambda)$ = $A_0 (R_\mathrm{p}(\lambda)/a)^2$ (assuming that the planet is seen at opposition), where $R_\mathrm{p}(\lambda)$ is the wavelength-dependent apparent planetary radius and $a$ is the semi-major axis.
As for the temperature structure, we use a free parametrization \citep{Pelletier_2021,Coulombe2023,Bazinet2024} which here fits for $N = 10$ temperature points ($\mathcal{U}[100,4500]$\,K) with fixed spacing in log-pressure (P = 10$^2$ -- 10$^{-8}$\,bar). Although this parametrization is free, it is regularized by a prior punishing for the second derivative of the profile using a physical hyperparameter, $\sigma_s$, with units of kelvin per pressure decade squared (K\,dex$^{-2}$) \citep{Pelletier_2021}. This prior is implemented to prevent nonphysical temperature oscillations at short pressure scale lengths. For this work, we use a hyperparameter value of $\sigma_s =$ 300\,K\,dex$^{-2}$ as we find this value to result in the best compromise between allowing flexibility of the temperature-pressure profile and avoiding over-fitting the observations.

Horizontal advection for hot tidally-locked gas giants is expected to homogenize atmospheric abundances to that of the dayside, resulting in the nightside being in chemical disequilibrium \citep{Agundez_2014,Showman_2020,Bell2024}. To account for this effect, we also run free chemistry retrievals where molecular and atomic abundances are kept constant with altitude. We fit for $\log$H$_2$O, $\log$CO, and $\log$H$^-$ ($\mathcal{U}[-16,0]$), as well as a cloud-top pressure $\log$P$_\text{cloud}$ and albedo $A$.
We find that the retrieved temperature and albedo distributions from the free chemistry retrievals are in good agreement with those measured from the chemical equilibrium results (see Extended Data Fig. \ref{fig:Teff_Ag_all}).

For the retrievals, we use four walkers per free parameter and consider the standard chi-square likelihood for the spectra fits. We run the retrieval for 30,000 steps and discard the first 18,000 steps, 60\% of the total amount, to ensure that the samples are taken after the walkers have converged. Spectra are initially computed using opacity sampling at a resolving power of R = 15,625, which is sufficient to simulate JWST observations \citep{Rocchetto_2016}, convolved to the instrument resolution and subsequently binned to the retrieved wavelength bins.

Because SCARLET assumes that the planet is seen at opposition when modeling the reflected light component of the planetary flux, we must correct for the orbital phase when interpreting our phase-resolved albedos. To do this, we first compute the measured reflected light planet-to-star flux ratio for a given orbital phase as $F_\mathrm{p,refl}/F_\mathrm{s} $ = $A_0 (R_\mathrm{p}/a)^2$, producing the values shown in Fig. \ref{fig:Ag_Teff_vs_phase}a. We then compute the phase-resolved albedos by dividing our measured $F_\mathrm{p,refl}/F_\mathrm{s}$ values by the integral of the visibility and illumination functions over the full surface (assuming Lambertian reflection), such that $A = (F_\mathrm{p,refl}/F_\mathrm{s})(a/R_\mathrm{p})^2/$($\frac{3}{2}\int_\Omega V I d\Omega$). The phase-resolved albedo values are shown in Fig. \ref{fig:Ag_Teff_vs_phase}c and Extended Data Fig. \ref{fig:Teff_Ag_all}. We measure albedos of 0.41$\pm$0.10, 0.50$\pm$0.07, and 0.79$\pm$0.15, at phases 120$^\circ$, 180$^\circ$, and 240$^\circ$, respectively, and find the albedo to be unconstrained for all phases spanning the nightside. Our dayside albedo measurement of 0.50$\pm$0.07 is consistent within 1.6$\sigma$ of the CHEOPS observations which found a geometric albedo of 0.80$_{-0.17}^{+0.10}$ (ref. \citep{Hoyer2023}). We note however that our measured albedo accounts only for the contribution from reflected light whereas the CHEOPS measurement corresponds to the sum of the reflected light and thermal emission components over its bandpass. Similarly to the reflected light component, we condense the information from the thermal portion of the measured planetary flux into a single value: the effective temperature $T_\mathrm{eff}$. To compute the effective temperature, we first begin by converting our measured thermal emission spectra into values of $F_\mathrm{p,therm}/F_\mathrm{s}$ where the planetary thermal emission, weighted by the throughput of NIRISS/SOSS and the PHOENIX stellar spectrum of LTT 9779, has been integrated from 1.0 to 2.8\,µm. The measured values of $F_\mathrm{p,therm}/F_\mathrm{s}$ for each orbital phase are shown in Fig. \ref{fig:Ag_Teff_vs_phase}b. To go from values of $F_\mathrm{p,therm}/F_\mathrm{s}$ to effective temperatures, we then compute the blackbody temperature that corresponds to the measured bandpass-integrated thermal emission. The phase-resolved values of effective temperatures are shown in Fig. \ref{fig:Ag_Teff_vs_phase}d and Extended Data Fig. \ref{fig:Teff_Ag_all}. We measure effective temperatures of 1,930$_{-60}^{+60}$\,K, 2,260$_{-50}^{+40}$\,K, and 1,850$_{-110}^{+100}$\,K, at phases 120$^\circ$, 180$^\circ$, and 240$^\circ$, respectively. Our dayside effective temperature is consistent with the measurement of $T_\mathrm{eff,day}$ = 2,305$\pm$141\,K obtained from the 3.6\,µm \textit{Spitzer}/IRAC secondary eclipses\citep{Dragomir2020}. We also measure 2-$\sigma$ upper limits on the effective temperature at phases 0$^\circ$, 60$^\circ$, and 300$^\circ$ of 1,171\,K, 1,202\,K, and, 1,452\,K, respectively, in agreement with the 2-$\sigma$ upper-limit of $T_\mathrm{eff,night}<$1,350\,K measured from the 4.5\,µm \textit{Spitzer}/IRAC phase curve\citep{Crossfield2020}. Furthermore, we find that the atmosphere has a non-inverted temperature-pressure profile for all dayside phases (Fig. \ref{fig:TP_vs_phase_row}), which is also consistent with refs.\citep{Dragomir2020,Crossfield2020} that found lower eclipse depths at 4.5\,µm compared to the 3.6\,µm channel, most likely due to CO/CO$_2$ absorption around 4.5\,µm.

Our measurements of the dayside and nightside effective temperatures of LTT\,9779b, obtained from spectroscopic observations that cover most of its spectral energy distribution, provide precise constraints on its energy budget and circulation regime.
Following the relations of ref.\citep{Cowan_2011} (eqs. 4 and 5), we derive a Bond albedo of $A_\mathrm{B}$ = 0.31$\pm$0.06 and a circulation efficiency of $\varepsilon<0.28$ (3-$\sigma$ confidence; $\varepsilon$ = 0 correspond to no redistribution and $\varepsilon$ = 1 to full redistribution). Combined with our reflected light geometric albedo measurement of $A_g$ = 0.50$\pm$0.07, we find that LTT\,9779b shows geometric and Bond albedos that are similar to that of Saturn ($A_g$ = 0.499, $A_\mathrm{B}$ = 0.342)\citep{Mallama_2017,Hanel1983}, Uranus ($A_g$ = 0.488, $A_\mathrm{B}$ = 0.300)\citep{Mallama_2017,Pearl1990}, and Neptune ($A_g$ = 0.442, $A_\mathrm{B}$ = 0.290)\citep{Mallama_2017,Pearl1991}.

We find that the cloud top pressure is unconstrained and spans the full prior range for all phases, most likely due to a degeneracy between the temperature-pressure profile and the cloud top pressure as observed in ref.\citep{Bell2024}. We also compare our measured short-wavelength dayside planet-to-star flux ratio spectrum to reflected light models produced with the \texttt{VIRGA}\citep{Rooney_2022} and \texttt{PICASO}\citep{Batalha_2019} python packages (Extended Data Fig. \ref{fig:cloud_models}). \texttt{VIRGA} takes as input a temperature-pressure profile, the mean molecular weight, as well as the atmospheric metallicity and recommends cloud species whose condensation curves cross the temperature-pressure profile. \texttt{PICASO} is then used to produce a reflected light spectrum for a given cloud species, sedimentation efficiency $f_\mathrm{sed}$, and vertical mixing coefficient $K_{zz}$.
For the temperature-pressure profile, we give as input a profile corresponding to the 25$^\text{th}$ percentile of the samples from the chemical equilibrium retrieval on the phase = 180$^\circ$ spectrum. The condensates recommended by \texttt{VIRGA} for a solar-metallicity (M/H = 1, $\mu$ = 2.2) atmosphere are Al$_2$O$_3$, Cr, Fe, Mg$_2$SiO$_4$, MgSiO$_3$, MnS, and TiO$_2$.
We use the 25$^\text{th}$ percentile of the temperature-pressure profile instead of the median as otherwise the only recommended species at these higher temperatures are Al$_2$O$_3$, Fe, MgSiO$_3$, and TiO$_2$, limiting the diversity of clouds that would be considered for this comparison. The reflected light models are then computed by assuming a sedimentation efficiency of $f_\mathrm{sed}$ = 0.1 and a vertical mixing coefficient of $K_{zz}$ = 10$^5$\,m$^2/$s, the same values used in the analysis of the CHEOPS observations of LTT\,9779b\citep{Hoyer2023}. We find that only silicate clouds (Mg$_2$SiO$_4$ and MgSiO$_3$) produce reflected light spectra that can explain the high albedo of LTT\,9779b, with all other species showing significantly lower planet-to-star flux ratios within the same wavelength range.

From the chemical equilibrium retrievals, the atmospheric metallicity and carbon-to-oxygen ratio are mostly unconstrained at all orbital phases, with a preference for low metallicities ([M/H] $<$ 1.54 at 3-$\sigma$ confidence) and high carbon-to-oxygen ratios (C/O = 0.871$_{-0.089}^{+0.041}$) at phase 180$^\circ$. Despite these two parameters being unconstrained, we find that the resulting water abundances have bounded constraints of $\log$H$_2$O = -4.51$_{-0.29}^{+0.39}$ and -4.68$_{-0.30}^{+0.37}$ at phases 120$^\circ$ and 180$^\circ$, respectively. From the free chemistry retrievals, we obtain a bounded constraint on the water abundance of $\log$H$_2$O = -5.04$_{-0.26}^{+0.25}$ at phase 180$^\circ$, consistent with the value from the chemical equilibrium retrieval, and find it to be unconstrained at all other phases. We note that the low metallicity and high C/O ratio measured from the chemical equilibrium retrieval at phase 180$^\circ$ might be mainly driven by the model trying to reduce the absolute water abundance, rather than by a constraint on the bulk composition or on any carbon-bearing species. These values are all below the solar value of $\log$H$_2$O = -3.21 (ref. \cite{Kreidberg_2014}).
Interpreting our measured water abundance of LTT\,9779b as a direct measurement of its atmospheric metallicity would correspond to a sub-solar metallicity, in apparent conflict with the mass-metallicity trend observed in solar system planets and transiting exoplanets \citep{Kreidberg_2014,Welbanks_2019}, as well as JWST NIRISS/SOSS transmission observations of this planet \citep{Radica2024}.
The condensation of clouds at high temperatures, which favors oxygen-rich species such as MgSiO$_3$, Mg$_2$SiO$_4$, Al$_2$O$_3$, and TiO$_2$, can deplete the atmospheric oxygen abundance by up to 30\% and by consequence increase the carbon-to-oxygen ratio\citep{Lee_2016,Lines2018,Pelletier_2021,grant2023jwsttst}. Considering a solar bulk atmosphere water abundance ($\log$H$_2$O = -3.21) for this planet, a 30\% depletion of gas-phase oxygen would result in a water abundance of $\log$H$_2$O = -3.36, which is not significant enough to explain our low measured H$_2$O abundance.
A possible explanation is that our abundance measurement is biased due to effects, such as spatial inhomogeneities across the dayside, suppressed molecular features due to clouds, as well as water dissociation, that are not accurately accounted for in our atmosphere model.
Nevertheless, if our measurement is accurate despite these caveats, the low bulk water abundance of LTT\,9779b could be the result of its formation history. Beyond the water ice line, the condensation of H$_2$O results in metal-poor and super-stellar C/O protoplanetary gas \citep{_berg_2011,Madhusudhan_2014}. Assuming that it formed beyond the water ice line via gas-dominated accretion, disk-free migration to its current-day location would preserve its primordial composition and explain our measured H$_2$O abundance. 
Thermal emission observations of LTT\,9779b using ground-based high-resolution spectroscopy or space-based observations at longer wavelengths could provide further constraints on its atmospheric C/O through measurement of its CO and CO$_2$ abundances, species to which JWST NIRISS/SOSS is less sensitive\citep{Coulombe2023,Taylor2023}.

\subsection{Energy Balance Models.}\label{sec:energy_balance}

We compute temperature maps from the energy balance model derived in ref.\citep{Crossfield2020} to compare with our measured phase-resolved thermal emission.
This model describes the temperature ($T'(\phi) = T/T_\mathrm{irr}$) at longitude $\phi$, normalized by the latitude-dependent irradiation temperature ($T_\mathrm{irr}(\theta) \equiv T_\mathrm{eff} [(1 - A_\mathrm{B}) \sin\theta]^{1/4}\sqrt{R_\mathrm{s}/a}$), as a function of the radiative-to-advective timescale ratio ($\varepsilon' \equiv \tau_\mathrm{rad}/\tau_\mathrm{adv}$):

\begin{equation}\label{eq:balance_mod}
T'(\phi) =
\bigg\{
    \begin{array}{lr}
        (1 - \varepsilon'^2/64\pi^2) + \varepsilon'\phi / 32\pi - \phi^2/8, & \text{if } -\pi/2\leq\phi\leq\pi/2 \\
        \left(6\pi\left[\phi-\pi/2\right]/\varepsilon'+[T'(\phi=\pi/2)]^{-3}\right)^{-1/3}, & \text{if } ~~~\pi/2<\phi< 3\pi/2
    \end{array}.
\end{equation}

\medskip
\noindent 
From this equation, we compute the temperature distribution ($T(\theta,\phi) = T'(\phi) T_\mathrm{irr}(\theta)$) for a grid of 18 pixels in latitude by 36 pixels in longitude. For the irradiation temperature, we assume a stellar effective temperature of $T_\mathrm{eff}$ = 5445\,K (ref.\citep{Jenkins2020}), and a ratio of the stellar radius to the semi-major axis of $R_\mathrm{s}/a$ = 0.259 (as measured from the white-light curve).
As for the Bond albedo, we use a value of $A_\mathrm{B}$ = 0.4 for the energy balance models as we find that this best reproduces our measured dayside effective temperature. This difference in the Bond albedo value that corresponds to the observed dayside temperatures between the energy balance model of ref.\citep{Crossfield2020} and the relations of ref.\citep{Cowan_2011} ($A_\mathrm{B}$ = 0.31$\pm$0.06) is most likely due to differences in the assumptions used to derive these models.
For each pixel of the grid, we convert the temperatures to values of planet-to-star flux ratio by assuming blackbody emission and integrating the flux from 1.0 to 2.8\,µm weighted by the throughput of NIRISS/SOSS and the PHOENIX model stellar spectrum of LTT 9779. We then transform our grid of $F_\mathrm{p}/F_\mathrm{s}$ values as a function of position on the planet to a phase curve by summing the product of the flux of each pixel to the thermal emission kernel of eq. \ref{eq:kernels}:

\begin{equation}\label{eq:pc_ebm}
F_\mathrm{p}(t) = \sum_{i=1}^{M=18}\sum_{j=1}^{N=36} \left[F_{i,j} \cdot \mathcal{K}_{\mathrm{therm},i,j}(t)\right] 
\end{equation}

\medskip 
\noindent 
where $i$ and $j$ correspond to the indices for latitude and longitude, respectively, and $F_{i,j}$ is the flux value of a given pixel $i,j$. We compute $F_\mathrm{p}(t)$ at 100 equally-spaced phases ranging from -180$^\circ$ to 180$^\circ$ for values of $\varepsilon'$ = 0.1, 1, 5, and 10, as shown in Figure \ref{fig:Ag_Teff_vs_phase}.

Cases where a significant amount of heat is re-emitted from the nightside ($\tau_\mathrm{adv}\gg\tau_\mathrm{rad}$) cannot explain our measured thermal emission distribution which is symmetric about the substellar point and shows low nightside flux. This is consistent with the case of a high-metallicity planet and/or a planet with nightside clouds. High atmospheric metallicity is known to dampen the advection of heat to the nightside due to the increase in opacity which leads to shorter radiative timescales as well as the increase in mean molecular weight which results in longer dynamical timescales\citep{Lewis2010,Zhang2017,Drummond_2018}. As for the scenario of a planet with a nightside that is covered in clouds, there can still be a significant advection of heat eastward of the substellar point and thus, an asymmetric temperature distribution across the dayside. However, the presence of a sharp gradient in flux near the terminators due to the clouds increases the phase curve amplitude and decreases the phase curve offset\citep{Parmentier_2020}. Furthermore, the presence of nightside clouds could result in a phase curve that shows a sharper decrease in flux away from the substellar point\citep{Parmentier_2020}, consistent with the fact that the energy balance models have difficulty reproducing our measured gradient in flux towards the nightside (Fig. \ref{fig:Ag_Teff_vs_phase}).



\end{methods}


\noindent\textbf{Data availability}\\
The data used in this work are publicly available in the Mikulski Archive for Space Telescopes ({\small \url{https://archive.stsci.edu/}}) under GTO program \#1201 (P.I. D. Lafrenière). The data used to create the figures in this manuscript are available on Zenodo (\url{https://doi.org/10.5281/zenodo.14232524}). All further data are available on request.

\noindent{\textbf{Code availability}}\\
The open-source codes that were used throughout this work are listed below:\\
\texttt{batman} (\url{https://github.com/lkreidberg/batman});\\
\texttt{emcee} (\url{https://emcee.readthedocs.io/en/stable/});\\
\texttt{celerite} (\url{https://celerite.readthedocs.io/en/stable/});\\
\texttt{VIRGA} (\url{https://natashabatalha.github.io/virga/});\\
\texttt{PICASO} (\url{https://natashabatalha.github.io/picaso/});\\

\begin{addendum}
\item This work is based on observations made with the NASA/ESA/CSA JWST. This project was undertaken with the financial support of the Canadian Space Agency (grant 18JWSTGTO2). The data were obtained from the Mikulski Archive for Space Telescopes at the Space Telescope Science Institute, which is operated by the Association of Universities for Research in Astronomy, Inc., under NASA contract NAS 5-03127. L.-P.C. acknowledges funding by the Technologies for Exo-Planetary Science (TEPS) Natural Sciences and Engineering Research Council of Canada (NSERC) CREATE Trainee Program. He would also like to thank Daniel Huber and Brett M. Morris for useful discussions on stellar granulation. B.B. and L.-P.C. acknowledge financial support from the Canadian Space Agency under grant 23JWGO2A05. M.R. acknowledges funding from NSERC. R.A.\ is a Trottier Postdoctoral Fellow and acknowledges support from the Trottier Family Foundation. D.J.\ is supported by NRC Canada and by an NSERC Discovery Grant. R.J.M. is supported by NASA through the NASA Hubble Fellowship grant HST-HF2-51513.001, awarded by the Space Telescope Science Institute, which is operated by the Association of Universities for Research in Astronomy, Inc., for NASA, under contract NAS 5-26555. J.F.R. acknowledges Canada Research Chair program and NSERC Discovery. J.D.T was supported for this work by NASA through the NASA Hubble Fellowship grant HST-HF2-51495.001-A awarded by the Space Telescope Science Institute, which is operated by the Association of Universities for Research in Astronomy, Incorporated, under NASA contract NAS5-26555. This research made use of the \texttt{Astropy}\citep{Astropy2022}, \texttt{Matplotlib}\citep{Hunter:2007}, \texttt{NumPy}\citep{harris2020array}, and \texttt{SciPy}\citep{2020SciPy-NMeth} python packages.
 
\subsection{Author contributions}
L.-P.C., M.R., and B.B. led the writing of this manuscript. D.L. provided the program leadership and designed the observational setup. L.-P.C. carried out the data reduction with contributions from M.R.. L.-P.C. performed the light curve fitting and atmospheric analysis. E.D. and L.-P.C. produced the clouds reflected light models. L.-P.C. produced the energy balance models. All co-authors provided significant comments and suggestions to the manuscript.

\item[Competing Interests] The authors declare that they have no competing financial interests.
 
\item[Correspondence] Correspondence and requests for materials should be addressed to Louis-Philippe Coulombe.~(email: louis-philippe.coulombe@umontreal.ca).
 
\end{addendum}



\clearpage
\setcounter{page}{1}
\setcounter{figure}{0}
\renewcommand{\figurename}{Extended Data Fig.}
\renewcommand{\tablename}{Extended Data Table}

\begin{center}
\textbf{\huge{}Extended Data}{\Huge\par}
\par\end{center}

\vspace{0mm}
\begin{figure}[H]
    \centering
    \includegraphics[width=1.\columnwidth]{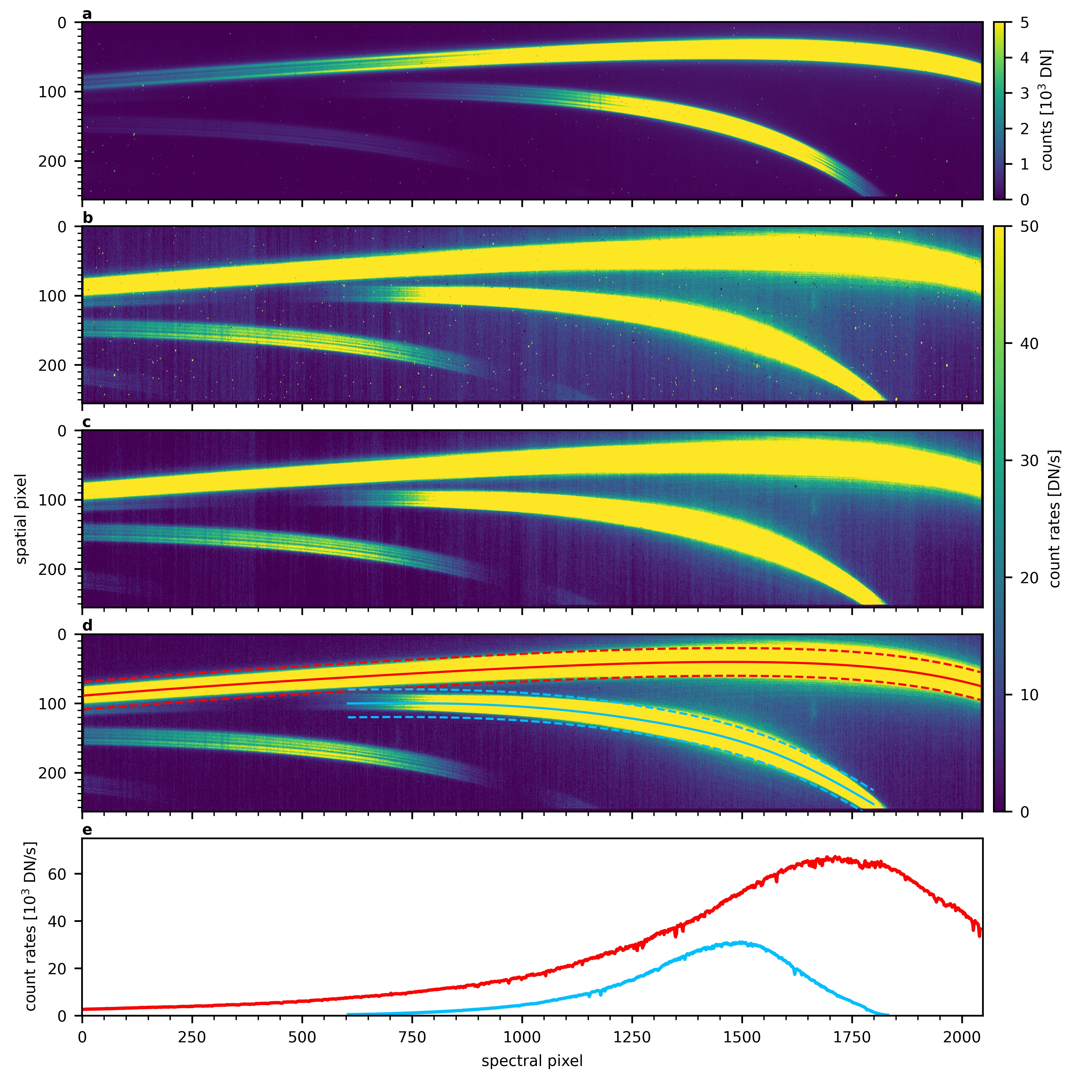}
    \caption{\sep \textbf{Reduction from the uncalibrated data to spectral extraction.} \textbf{a,} Last group of the 10$^{\text{th}}$ integration after subtraction of the super-bias and non-linearity correction. \textbf{b,} Frame of the 10$^\text{th}$ integration after ramp-fitting, converting the data from counts to count rates. \textbf{c,} Frame after 2D interpolation over the bad pixels, cosmic ray hits correction, and subtraction of the non-uniform background. \textbf{d,} Frame after correction of the $1/f$ noise, residual striping is still visible away from the first and second order traces due to intra-column variations of the noise. The extraction boxes with aperture widths of 40 pixels used for the first and second order are shown in red and blue, respectively. \textbf{e,} Spectra extracted from the frame. The first and second order spectra are shown in red and blue, respectively.}
    \label{fig:reduc_steps}
\end{figure}

\vspace{-0mm}
\begin{figure}[H]
    \centering
    \includegraphics[width=1.\columnwidth]{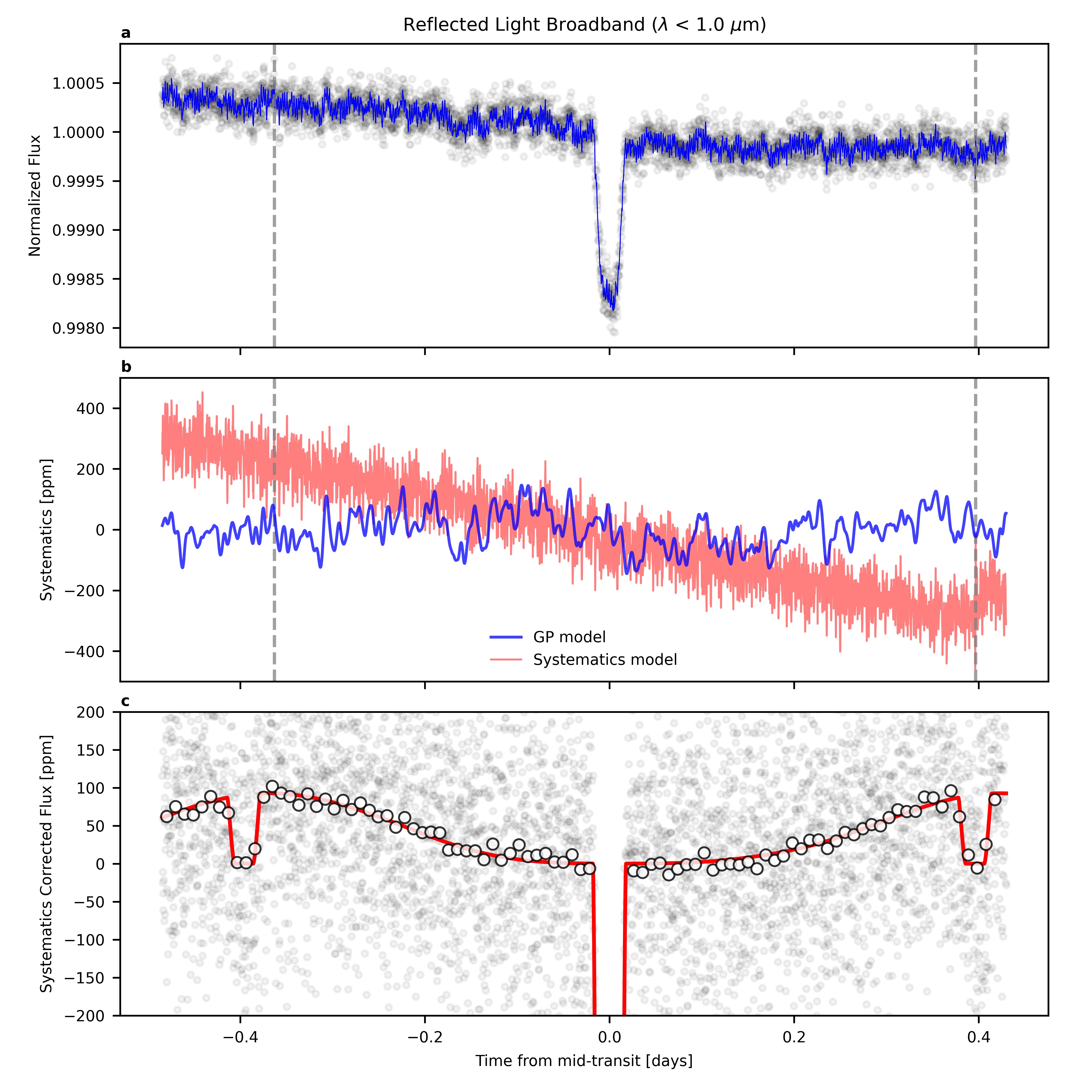}
    \caption{\sep \textbf{NIRISS/SOSS LTT\,9779b reflected light broadband light curve fit.} \textbf{a,} Raw white-light curve produced from the data with wavelengths $\lambda<$ 1\,µm. The data is shown by the black points and the fit, including the systematics and GP models, is shown in blue. The positions of the two tilt events are indicated by the gray dashed lines. \textbf{b,} Amplitude in ppm of the systematics and GP models removed from the data. \textbf{c,} Systematics-corrected white-light curve. The data is shown by the black points at the original temporal resolution and also at a resolution of 50 integrations per bin. The astrophysical model is shown in red.
    }
    \label{fig:wlc_fits}
\end{figure}

\vspace{-0mm}
\begin{figure}[H]
    \centering
    \includegraphics[width=1.\columnwidth]{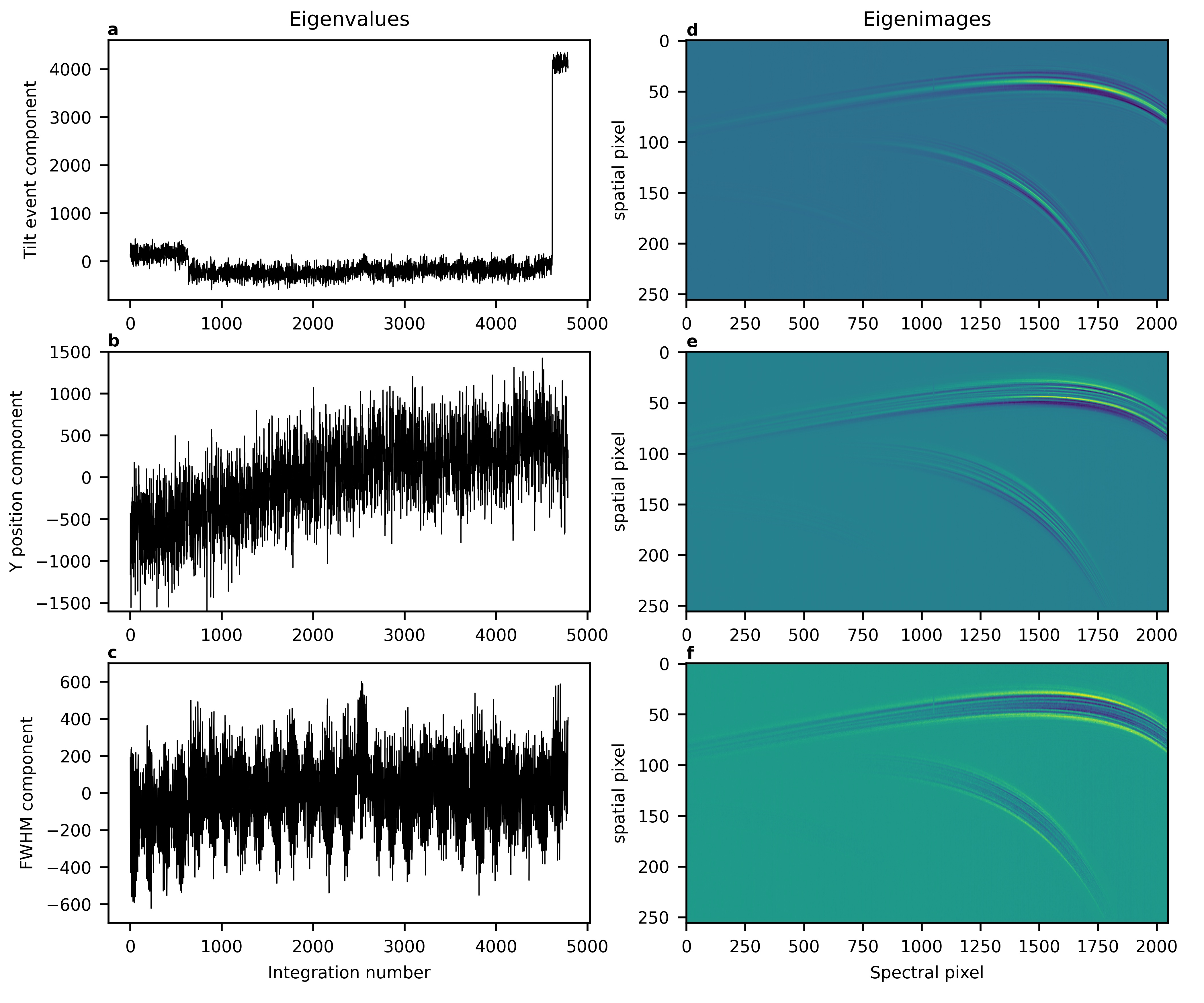}
    \caption{\sep \textbf{Principal component analysis of the detector time series.} Eigenvalues (\textbf{a, b,} and \textbf{c,}) and eigenimages (\textbf{d, e,} and \textbf{f,}) of the components that contribute most to the variance of the detector time series. \textbf{a,} The first component strongly picks up the two tilt events that occur at  the \nth{636} and \nth{4615} integrations. The tilt events result in changes in the overall trace morphology, as seen in the corresponding eigenimage (\textbf{d,}). \textbf{b,} The second component is analogous to the y position (spatial direction) of the spectral orders, as seen by the trade in flux between the two wings of the trace in its eigenimage (\textbf{e,}). \textbf{c,} The third component is analogous to the full width at half maximum (FWHM), which appears as a change in flux between the edges and center of the trace in its eigenimage (\textbf{f,}).}
    \label{fig:PCA}
\end{figure}

\begin{figure}[H]
    \centering
    \includegraphics[width=1.\columnwidth]{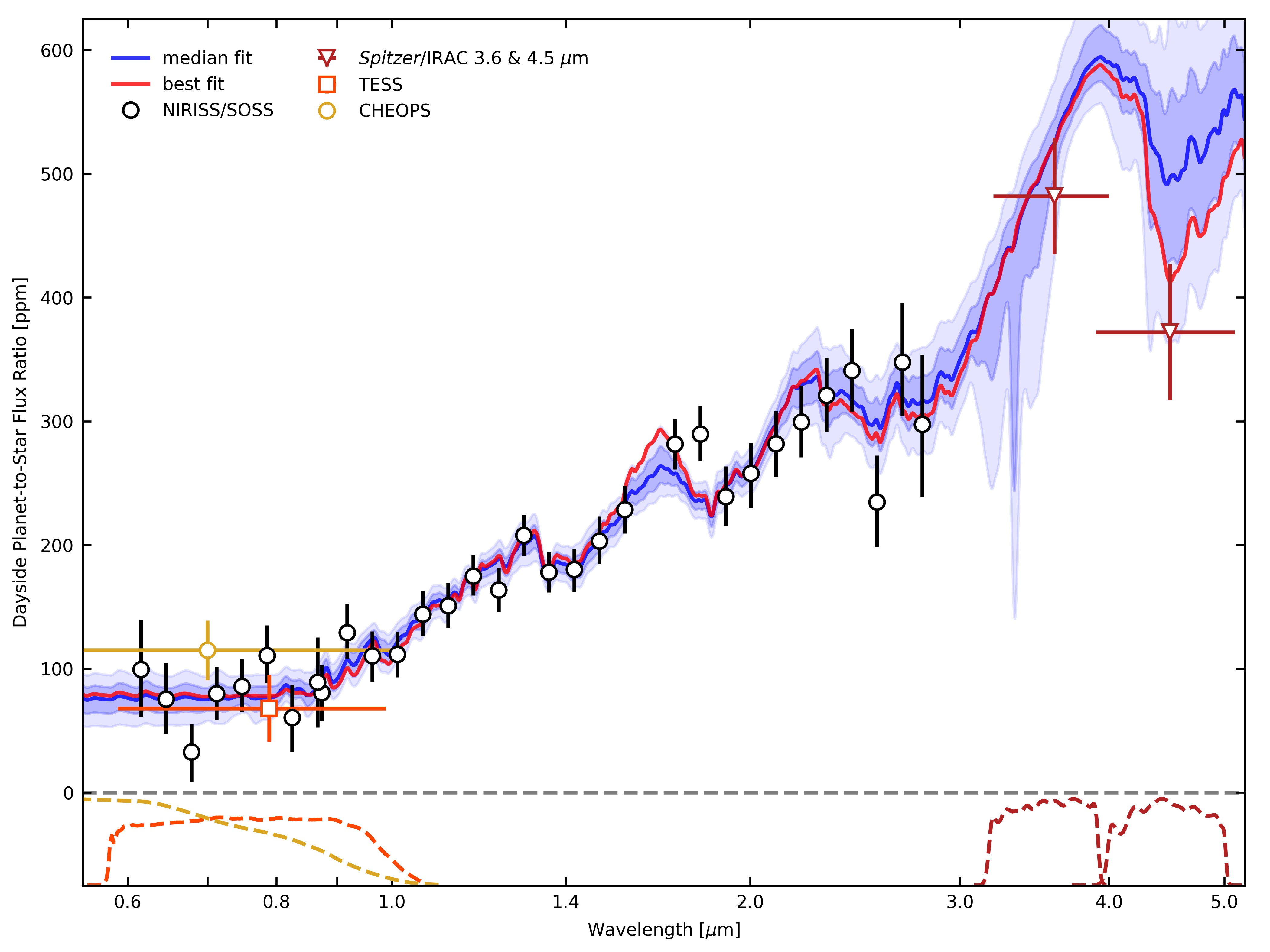}
    \caption{\sep \textbf{Dayside planet-to-star flux ratio spectrum of LTT 9779b.} Measured NIRISS/SOSS planet-to-star flux ratio at phase 180$^\circ$ (black points), shown along with the CHEOPS\citep{Hoyer2023} (yellow point), TESS (orange point), and \textit{Spitzer}/IRAC\citep{Dragomir2020} (red points) secondary eclipse measurements. The best fit retrieved spectrum from the chemical equilibrium retrieval to the NIRISS/SOSS data, extended to cover the wavelengths of past observations, is shown in red. The median retrieved spectrum is shown in blue, along with the 1 and 2-$\sigma$ confidence regions. Our measured reflected light spectrum at short wavelengths is consistent with the CHEOPS and TESS photometric points, and our measured thermal emission matches the \textit{Spitzer} 3.6 and 4.5 µm secondary eclipse measurements. The instrument response of the CHEOPS, TESS, and \textit{Spitzer} points are shown by the dashed lines.}
    \label{fig:sec_ecl_spec_ppm}
\end{figure}

\begin{figure}[H]
    \centering
    \vspace{-10mm}
    \includegraphics[width=1.\columnwidth]{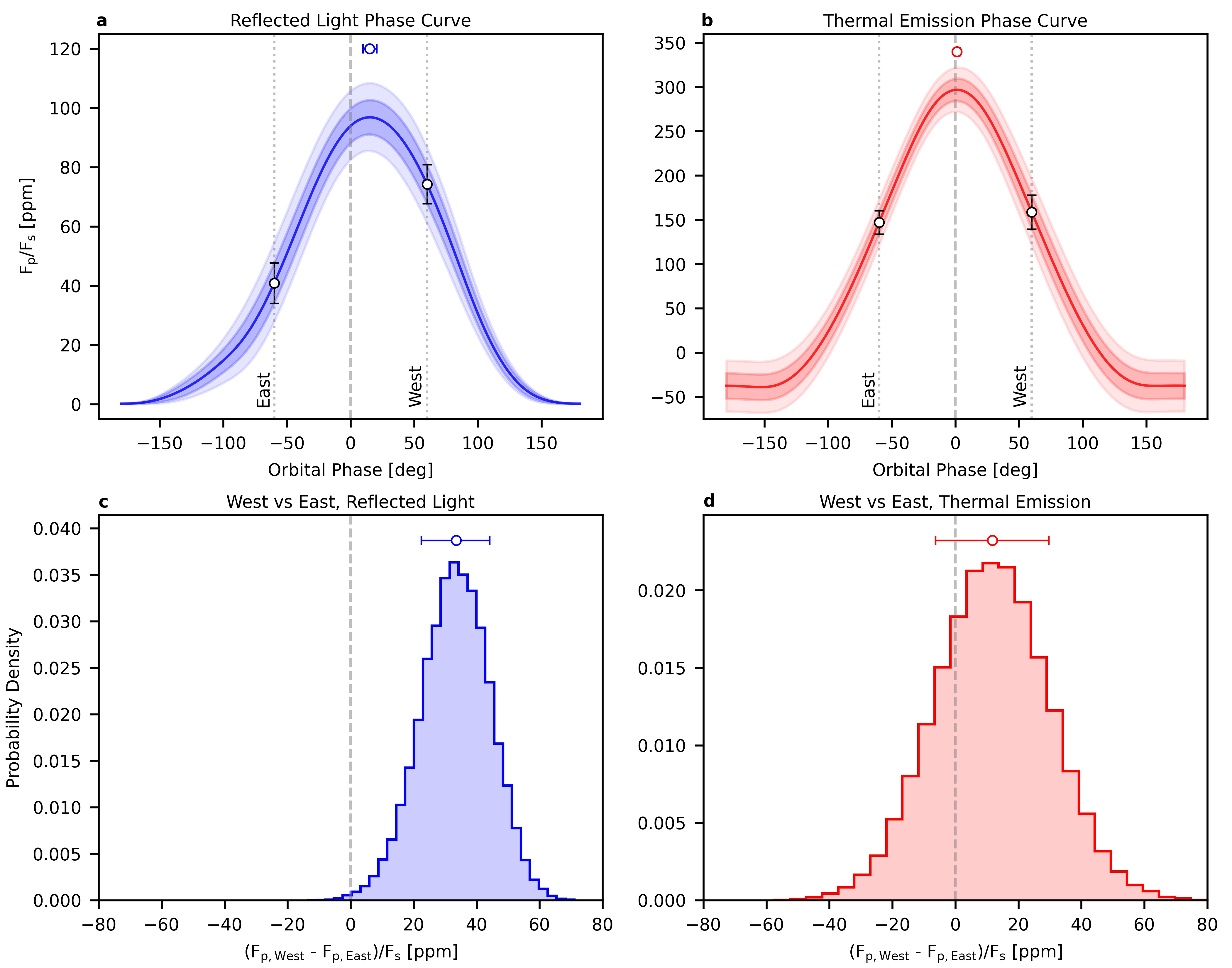}
    \caption{\sep \textbf{Measurement of an asymmetric dayside in reflected light and a symmetric phase curve in thermal emission. a,} Phase curve measured by fitting six albedo slices to the light curve produced by summing all wavelengths below 0.85\,µm (see Methods). The median phase curve model is shown by the blue line, along with the 1- and 2-$\sigma$ confidence intervals. The measured phase curve offset and its 1-$\sigma$ uncertainty is indicated by the blue dot, where the gray dashed line corresponds to a phase curve offset of 0. The phase curve peaks after secondary eclipse, corresponding to a westward phase curve offset. The planetary flux and their 1-$\sigma$ uncertainties at phases $\pm$60$^\circ$ are shown by the black dots. \textbf{b,} Phase curve measured by fitting six thermal emission slices to the light curve produced by summing all wavelengths above 1.9\,µm. The median phase curve model is shown by the red line, along with the 1- and 2-$\sigma$ confidence intervals. The measured phase curve offset and its 1-$\sigma$ uncertainty is indicated by the red dot. The phase curve peaks at secondary eclipse, which is indicated by the gray dashed line. The planetary flux and their 1-$\sigma$ uncertainties at phases $\pm$60$^\circ$ are shown by the black dots. \textbf{c,} Posterior probability distribution of the difference between the western dayside flux (phase = 60$^\circ$) and the eastern dayside flux (phase = -60$^\circ$) for the reflected light phase curve. We find that the measured difference is more than 3-$\sigma$ away from 0. The blue dot indicates the median and 1-$\sigma$ confidence interval of the posterior distribution. \textbf{d,} Posterior probability distribution of the difference between the western dayside flux (phase = 60$^\circ$) and the eastern dayside flux (phase = -60$^\circ$) for the thermal emission phase curve. We find that the measured difference is within 1-$\sigma$ of 0. The red dot indicates the median and 1-$\sigma$ confidence interval of the posterior distribution.}
    \label{fig:Aw_vs_Ae}
\end{figure}

\vspace{-0mm}
\begin{figure}[H]
    \centering
    \includegraphics[width=1.\columnwidth]{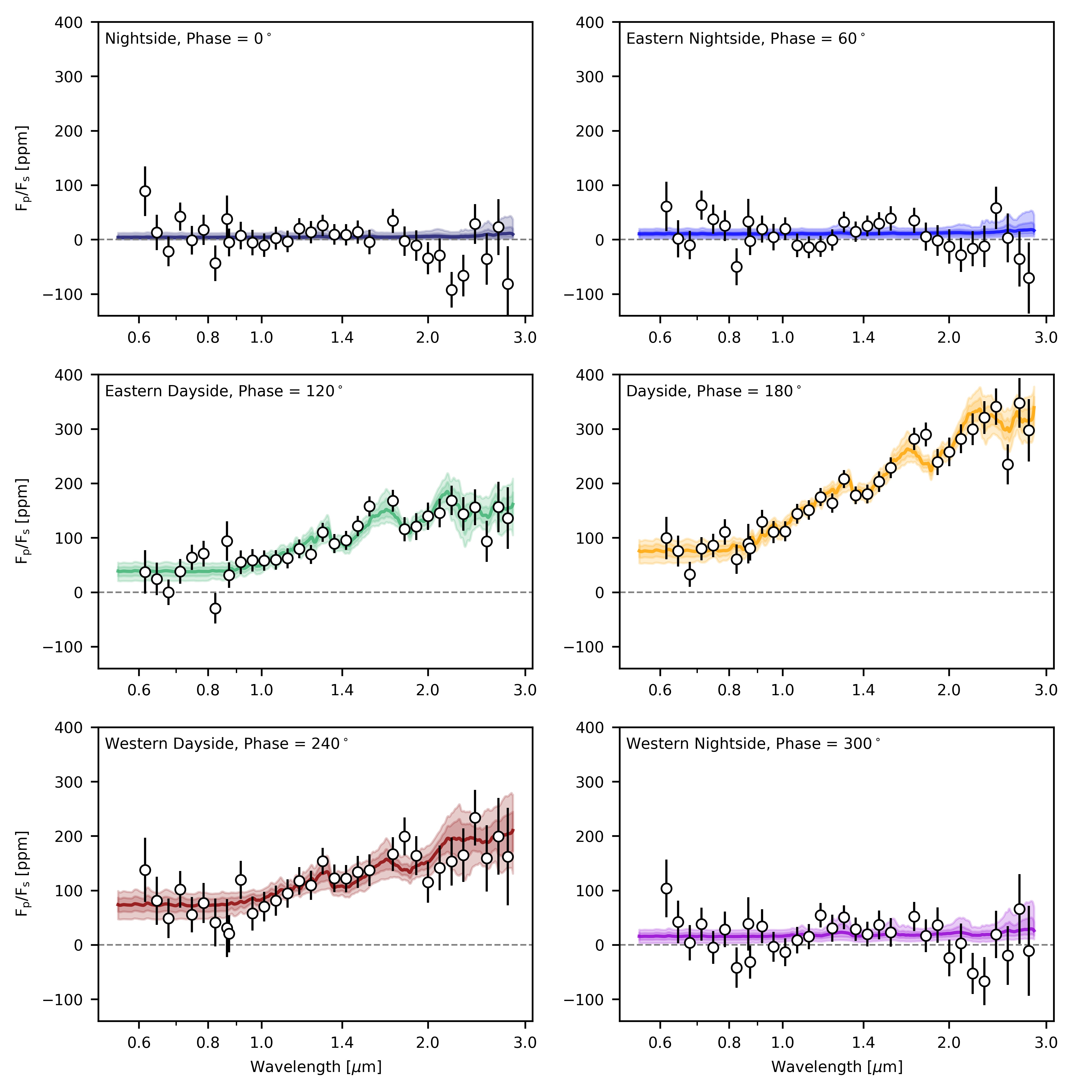}
    \caption{\sep \textbf{Phase-resolved planet-to-star flux ratio spectra of LTT\,9779b.} Measured NIRISS/SOSS emission spectra (black dots) at six orbital phases across the orbit of LTT\,9779b. The median retrieved models from the SCARLET chemical equilibrium retrievals are shown along with their 1 and 2-$\sigma$ confidence intervals. The atmospheric models all show water absorption features at phases 120$^\circ$, 180$^\circ$, and 240$^\circ$. The reflected light component, which dominates the planetary flux for wavelengths below 1\,µm, is noticeably higher on the western dayside compared to the eastern dayside. The retrieved reflected light and thermal emission components for phases 0$^\circ$, 60$^\circ$, and 300$^\circ$ are all upper-limits due to the low nightside temperatures of LTT\,9779b and the small amount of reflected light that reaches the observer at these orbital phases.}
    \label{fig:phase_specs}
\end{figure}


\begin{figure}[H]
    \centering
    \includegraphics[width=1.\columnwidth]{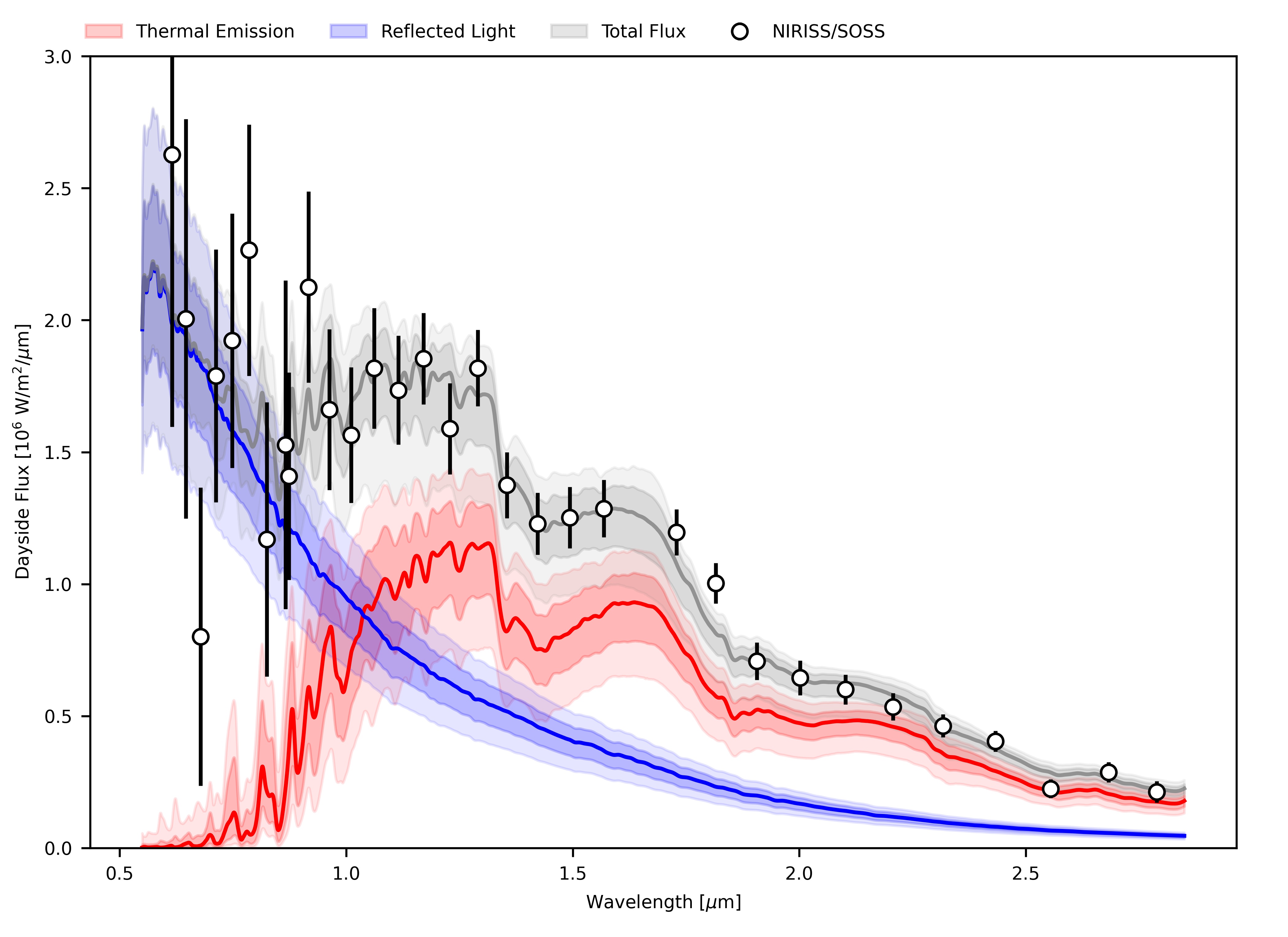}
    \caption{\sep \textbf{Constraints on the thermal and reflected light components of the planetary spectrum.} Measured dayside flux spectrum of LTT\,9779b in physical units (black points). The median retrieved model from the SCARLET chemical equilibrium retrieval at phase 180$^\circ$ (gray line) is shown along with its 1 and 2-$\sigma$ confidence intervals. The individual contributions of the reflected light and thermal emission to the total flux are shown in blue and red, respectively. The transition where thermal emission becomes dominant occurs around 1\,µm, which is consistent with the position of the observed transitions from reflected light to thermal emission in the spectra of the phase curve amplitudes (Fig. \ref{fig:spec_lcs}b) and offsets (Fig. \ref{fig:spec_lcs}c). The thermal emission captured over the wavelength range of NIRISS/SOSS corresponds to 80\% of the total thermal emission of the planet.}
    \label{fig:therm_refl_constraint}
\end{figure}

\begin{figure}[H]
    \centering
    \includegraphics[width=1.\columnwidth]{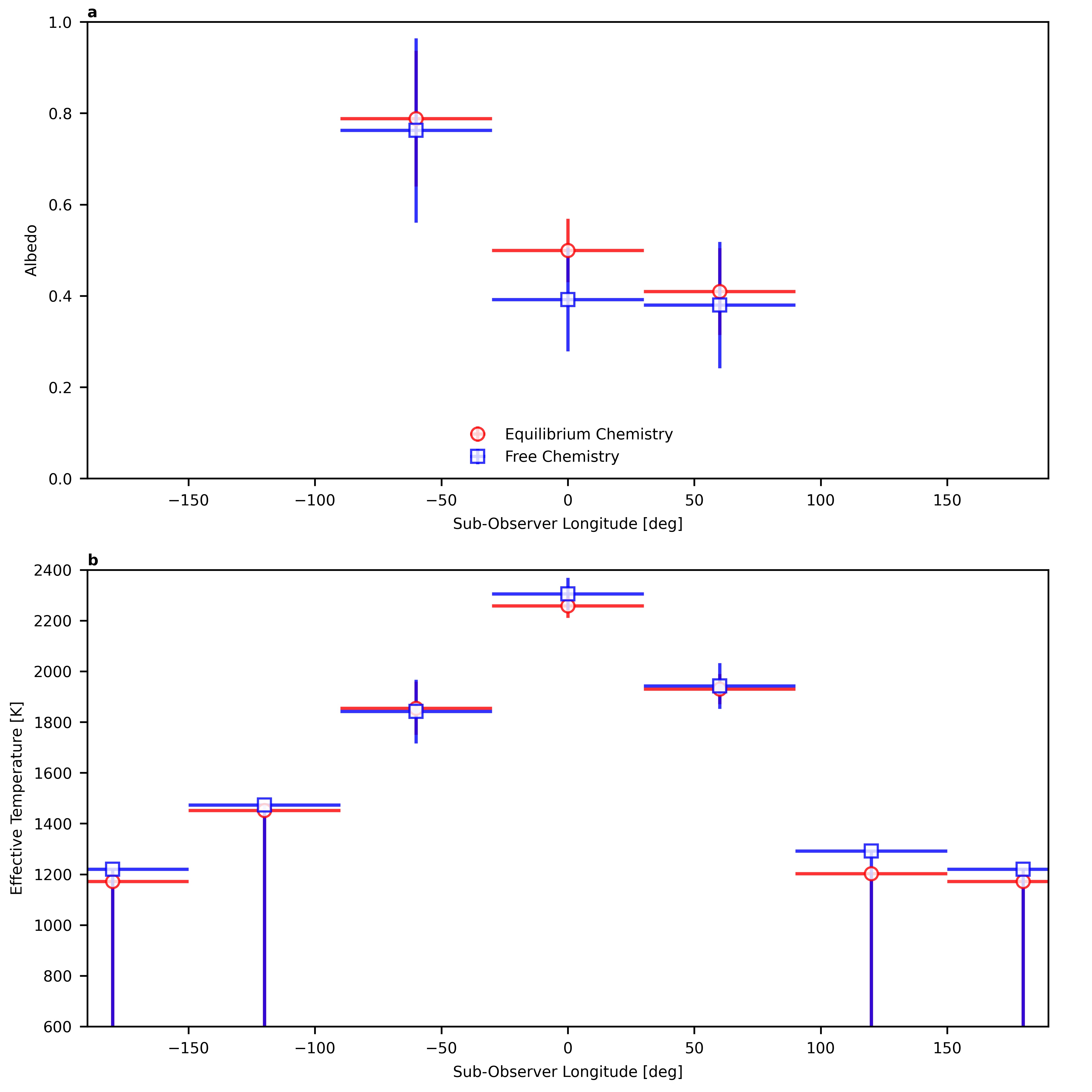}
    \caption{\sep \textbf{Impact of modeling assumptions on the retrieved albedo and effective temperature distributions. a,} Retrieved albedo values as a function of the sub-observer longitude from the chemical equilibrium (red dots) and free chemistry (blue squares) retrievals. The albedo values retrieved from both sets of retrievals agree at less than 1-$\sigma$ at all orbital phases and show the same east-west asymmetry in the albedo distribution. We do not show the albedo values past the terminators, as they are unconstrained. \textbf{b,} Retrieved effective temperature as a function of sub-observer longitude. As with the albedo values, the chemical equilibrium and free chemistry retrievals find effective temperatures that agree at less than 1-$\sigma$ at all orbital phases and both show a symmetric temperature distribution. The 2-$\sigma$ upper-limits are shown for the effective temperature values past the terminators.}
    \label{fig:Teff_Ag_all}
\end{figure}

\begin{figure}[H]
    \centering
    \includegraphics[width=1.\columnwidth]{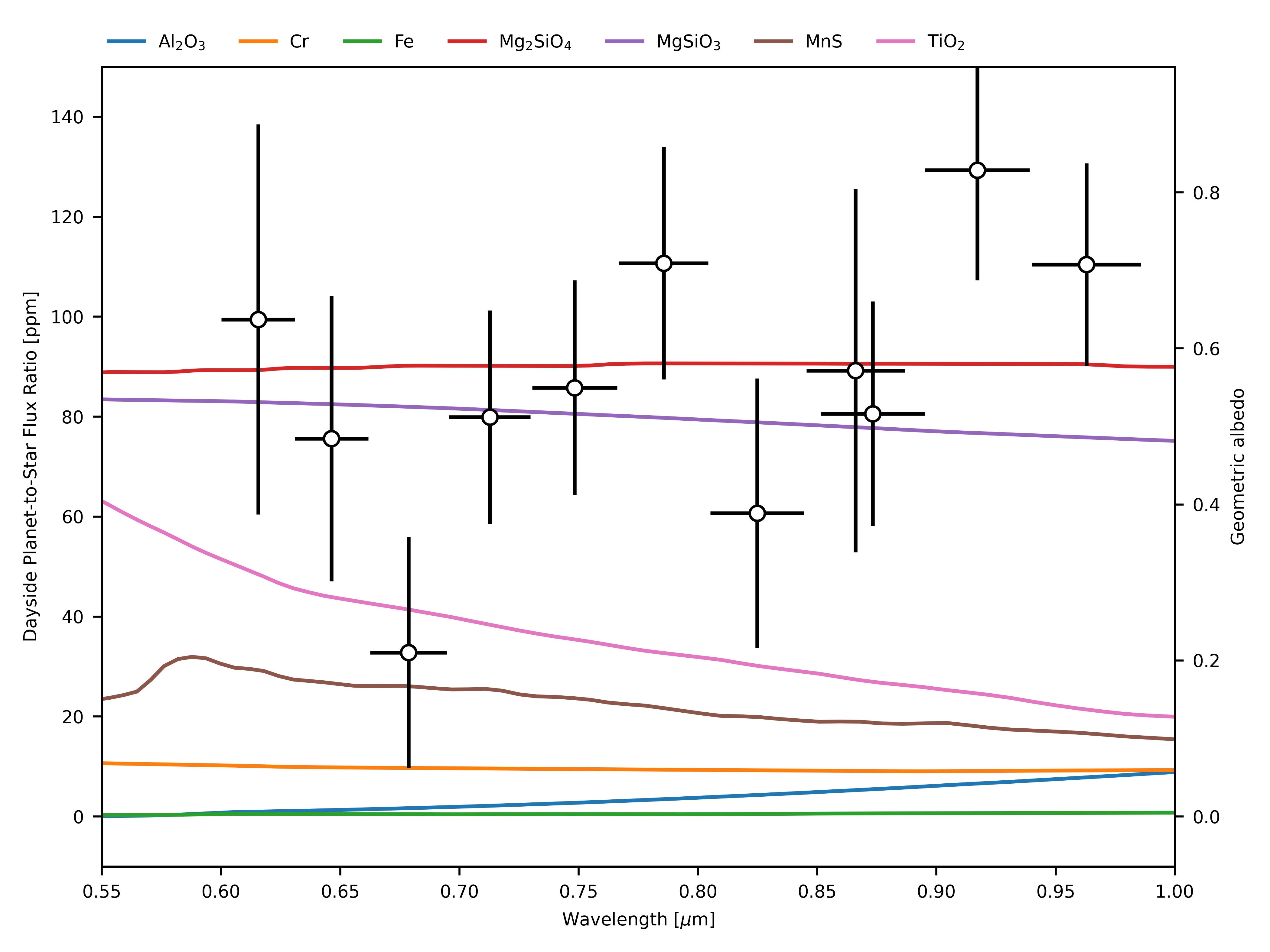}
    \caption{\sep \textbf{Wavelength-dependent geometric albedo of LTT\,9779b for various cloud species.} Reflected light models produced with \texttt{VIRGA}\cite{Rooney_2022} and \texttt{PICASO}\cite{Batalha_2019} for different cloud species that are expected to condense in the atmosphere of LTT\,9779b in units of planet-to-star flux ratio [ppm] and geometric albedo (see Methods). The NIRISS/SOSS (black points) occultation depths and their 1$\sigma$ error bars at phase 180$^\circ$ are shown for wavelengths below 1\,µm, where the contribution from thermal emission is negligible. The two cloud species that can produce similar geometric albedos are Mg$_2$SiO$_4$ and MgSiO$_3$.}
    \label{fig:cloud_models}
\end{figure}

\clearpage





\bibliography{main}

\end{document}